\documentclass[journal]{IEEEtran}
\renewcommand{\baselinestretch}{1.0}
\renewcommand{\theenumii}{\theenumi.\arabic{enumii}}

\usepackage[utf8]{inputenc}
\usepackage{amssymb}
\usepackage{color,soul}
\usepackage[T1]{fontenc}
\usepackage{multicol}
\usepackage{multirow}
\usepackage{textcomp}
\usepackage{multirow}
\usepackage{bigdelim}
\usepackage{graphicx}
\usepackage{epstopdf}
\usepackage{amssymb}

\usepackage{amsmath}
\usepackage{psfrag}
\usepackage{amsthm}
\usepackage[tight,footnotesize]{subfigure}
\usepackage{algorithmic,algorithm}
\usepackage{mathrsfs}
\usepackage{framed}
\usepackage{color}
\usepackage{multirow,enumerate}
\usepackage{bigdelim}
\definecolor{shadecolor}{rgb}{0.92,0.92,0.92}
\usepackage{soul}
\usepackage{color}
\usepackage{cuted}
\usepackage{cite}
\usepackage{booktabs}

\theoremstyle{definition}

\newtheorem{theorem}{Theorem}

\newtheorem{remark}{Remark}

\makeatletter
\newcommand{\vast}{\bBigg@{3.2}}
\newcommand{\Vast}{\bBigg@{4.5}}

\makeatother

\graphicspath{{Figure/}}
\begin{document}

\title{Global Iterative Sliding Mode Control of an Industrial Biaxial Gantry System for\\ Contouring Motion Tasks}

\author{
 Wenxin Wang, Jun Ma, Zilong Cheng, Xiaocong Li,
	\\
	Clarence W de Silva, \IEEEmembership{Fellow,~IEEE,} and
	Tong Heng Lee
	
\thanks{W. Wang, Z. Cheng, and T. H. Lee are with the Department of Electrical and Computer Engineering, National University of Singapore, Singapore 117583 (e-mail: wenxin.wang@u.nus.edu; zilongcheng@u.nus.edu; eleleeth@nus.edu.sg).}
\thanks{J. Ma, X. Li is with the John A. Paulson School of Engineering and Applied Sciences, Harvard University, Cambridge, MA 02138 USA (e-mail: junma@seas.harvard.edu; xiaocongli@seas.harvard.edu).}
\thanks{Clarence W. de Silva is with the Department of Mechanical Engineering, University of British Columbia, Vancouver, BC, Canada V6T 1Z4 (e-mail: desilva@mech.ubc.ca).}
\thanks{This work has been submitted to the IEEE for possible publication. Copyright may be transferred without notice, after which this version may no longer be accessible.}
}

\maketitle

\begin{abstract}

This paper proposes a global iterative sliding mode control approach for high-precision contouring tasks of a flexure-linked biaxial gantry system. For such high-precision contouring tasks, it is the typical situation that the involved multi-axis cooperation is one of the most challenging problems. As also would be inevitably encountered, various factors render the multi-axis cooperation rather difficult; such as the strong coupling (which naturally brings nonlinearity) between different axes due to its mechanical structure, the backlash and deadzone caused by the friction, and the difficulties in system identification, etc. To overcome the above-mentioned issues, this work investigates an intelligent model-free contouring control method for such a multi-axis motion stage. Essentially in the methodology developed here, it is firstly ensured that all the coupling, friction, nonlinearity, and disturbance (regarded as uncertain dynamics in each axis) are suitably posed as `uncertainties'. Then, a varying-gain sliding mode control method is proposed to adaptively compensate for the matched unknown dynamics in the time domain, while an iterative learning law is applied to suppress the undesirable effects (arising from the repetitive matched and unmatched uncertainties in the iteration domain). With this approach, the chattering that typically results from the overestimated control gains in the sliding mode control is thus suppressed during the iterations. To analyze the contouring performance and show the improved outcomes, rigorous proof is furnished on both the stability in the time domain and the convergence in the iteration domain; and the real-time experiments also illustrate that the requirements of precision motion control towards high-speed and complex-curvature references can be satisfied using the proposed method, without prior knowledge of the boundary to the unknown dynamics.

\end{abstract}


\section{Introduction}

An industrial biaxial gantry stage is a high-speed and high-precision planar robot system, which has been extensively used in different automated processes when precise Cartesian motion is required. Representative applications of such a biaxial gantry system include precision lithography, CNC machining, CT scanning, etc. Typically, the gantry system consists of two parallel carriages linked by an orthogonal cross-arm, and with an end-effector laden payload. Each carriage is driven by a linear motor, for which its characteristics of simple transmission and quick motion response are important underpinning characteristics~\cite{wang2020prediction}. Thus, as an instance of such an application in this class, the conventional X-Y table is certainly widely used in planar contouring tasks (such as scanning and shaping etc). In the more conventional configuration, the carriages and the cross-arm are linked by rigid joints, and this conventional design has demonstrated its effectiveness in minimizing various de-synchronization issues~\cite{yuan2016time,mobayen2017composite1,mobayen2018synchronization2,hu2020gru}. However, in certain scenarios when the involved de-synchronization requirements substantially increases to a higher degree, the large inter-axis coupling forces may damage the joints~\cite{ma2017novel}. To prevent this kind of unintended damage, alternative flexure-based mechanisms have been designed and extensively adopted instead in various appropriate classes of precision motion stages~\cite{xu2013design,zhu2016tmech,wu2018design,kang2020six}. Hence, such alternate gantry system designs replace the rigid joints with flexure joints, which (in such instances) enable a small degree of rotation of the cross-arm to prevent the afore-mentioned possibility of machine-wide damage; and thus a rather more cost-effective repair can be realized simply through joint replacement only when the unintended damages occur~\cite{ma2017integrated,kamaldin2018novel}. However, one of the drawbacks of this alternate flexure-based mechanism design is that it introduces more coupling and nonlinearity to the system; and which naturally also brings various uncertainties and disturbances during the contouring tasks.

To ensure the attainment of stringent contouring performance of the biaxial gantry stage, several approaches have been proposed for multi-axis cooperation control. One of the commonly used approaches is called cross-coupled control~\cite{yang2010novel}, where decoupling methods are applied to the multi-axis system so that each axis can be controlled based on its decoupled model respectively; and the resulting contouring problem can then be formulated using the methodology of the tracking problem therein~\cite{yepes2013evaluation}. However, it is pertinent to note that the decoupled model is essentially an approximation, and thus could limit the contouring performance attainable. As alternatives in the literature, the multi-axis coupling can also be modeled with certain assumptions and then a decentralized control scheme is developed for the gantry system~\cite{ma2018robust,ma2019parameter}. However, the system identification for such multi-axis systems is inevitably inaccurate~\cite{li2017data,li2020data}. With regards to the issues of multi-axis cooperation, the methodology of coordinated frames is also further proposed in the literature; such as the global task coordinate frame (GTCF)~\cite{yao2011orthogonal,hu2011global} and the Newton-ILC contouring control method~\cite{wang2017newton}. These contouring control methodologies can be suitably used for a rigid-linked biaxial gantry system. But unfortunately it is not straightforward to be extended for control of the flexure-linked counterpart, due to the inaccurate and unmeasured position of the cross-arm (the sensor is implemented in each carriage). Additionally, it is also noted in the literature that the robust controller methodology is also an appropriate option for contouring control of such systems; and thus for instance, some adaptive robust controllers are considered in the works reported in~\cite{chen2014mu,hu2016advanced,han2020time,hu2020precision,chen2020optimal}, though it needs to be noted that the identification of various system parameters is still required for initialization.

In such applications, it may also be noted that sliding mode control is one of the rather more preferred and effective nonlinear control methodologies, where the ideal sliding motion makes the overall system suitably insensitive to matched uncertainties/perturbations \cite{edwards1998sliding,xu2011micro,xu2012identification,li2021simultaneous}. Thus, good robustness against such uncertainties/perturbations~\cite{mobayen2016novel4} and nonlinearities~\cite{mobayen2018adaptive3} can be well guaranteed. Though for sliding mode control, it is well-known that two essential issues need consideration, \emph{i.e.,} the requirements of the knowledge of a bound characterizing the uncertainties (even though it is only the bound that needs to be known)~\cite{plestan2010new} and the need to properly handle the effects of chattering~\cite{lee2007chattering}. To suitably eliminate the need of the prior knowledge (of the uncertainty bound), the approach of varying-gain sliding mode control~\cite{huang2008adaptive,gonzalez2011variable} has been proposed to keep the control gain increasing until the sliding variables are driven and constrained within the sliding hyperplane. In such approaches however, the control gains will increase with a smaller rate as the sliding variables evolve closer to the sliding hyperplane; and thus the overall performance could essentially be degraded. Additionally, because the ideal sliding hyperplane is not reachable in practice, the control gains will be inevitably overestimated, which brings unnecessary chattering to the system. Thus to further solve this problem, a modified adaptive sliding mode control method is proposed in~\cite{shtessel2012novel}; but unfortunately here, the performance is then essentially sacrificed as a trade-off to attain the chattering suppression. In addition, it is pertinent to note that sliding mode control can typically only deal with matched uncertainties~\cite{wang2020iterative}, but will be unable to address any presence of unmatched uncertainties.

Taking all the above matters into consideration, a new formulation of a global iterative sliding mode control method is proposed in this paper. The method here utilizes a 2-degree-of-freedom (DOF) control structure, and is composed of an adaptive sliding mode controller with an incremental cascade iterative learning law. In this work, it is ensured that all of the nonlinearity, coupling (effects from other axes and other actuators), matched disturbances, and the uncertain knowledge for each carriage are suitably posed as `uncertainties'; and the formulation here then enables each carriage to track its own reference profile accurately to achieve the desired high-precision contouring performance while taking into account these uncertainties. Moreover, the establishment of sliding motion will be ensured in both the time domain and the iteration domain globally. Specifically, in this new formulation of a global iterative sliding mode control method, the first part of the composition involving the adaptive sliding mode control is designed to ensure the establishment of the sliding motion in finite time in each iteration; and next, the second part of the composition involving the iterative learning law~\cite{bristow2006survey} is designed to compensate for the repetitive uncertainties (not only matched uncertainties but also unmatched uncertainties that cannot be suppressed by previous works in the adaptive sliding mode control) in the iteration domain. As a result, the sliding hyperplane is made more effective and reachable in its attraction domain. Furthermore, the iterative learning law developed makes the procedure utilized here of a varying gain strategy incorporated into the iteration domain closely, so that the tracking performance and sliding precision can be significantly improved with typically lower control gain values (instead of conservative overestimation) and lower chattering (at least at the same levels and typically better and improved). The main contributions of this paper are thus as follows: (1). This work presents a totally model-free controller without any prior knowledge of the system parameters. Additionally, utilizing the methodology, the contouring performance can be guaranteed at a high level despite the existence of matched or unmatched uncertainties. (2). The uncertainties arising from matched unknown dynamics in the time domain are effectively compensated for (via the first part of the composition involving the adaptive sliding mode control); and the undesirable effects arising from the remaining repetitive matched and unmatched uncertainties in the iteration domain are also suppressed (via the second part of the composition involving the iterative learning law). (3). With the proposed control method, the sliding hyperplane is more attractive for the sliding variable, and the sliding accuracy is improved significantly without control gain overestimation and high chattering during the iterations.

The remainder of this paper is organized as follows. In Section II, the model of the biaxial gantry system and related issues in contouring tasks are provided and discussed. In Section III, the model-free control scheme developed and proposed in the work is presented. Also, rigorous proofs on convergence and stability are given. In Section IV, several real-time experiments are presented with detailed analysis. Finally, pertinent conclusions are drawn in Section V.

\section{Problem Formulation}
\begin{figure}[t]
	\centerline{\includegraphics[trim = 0cm 0cm 0cm 0cm, width=0.8\columnwidth]{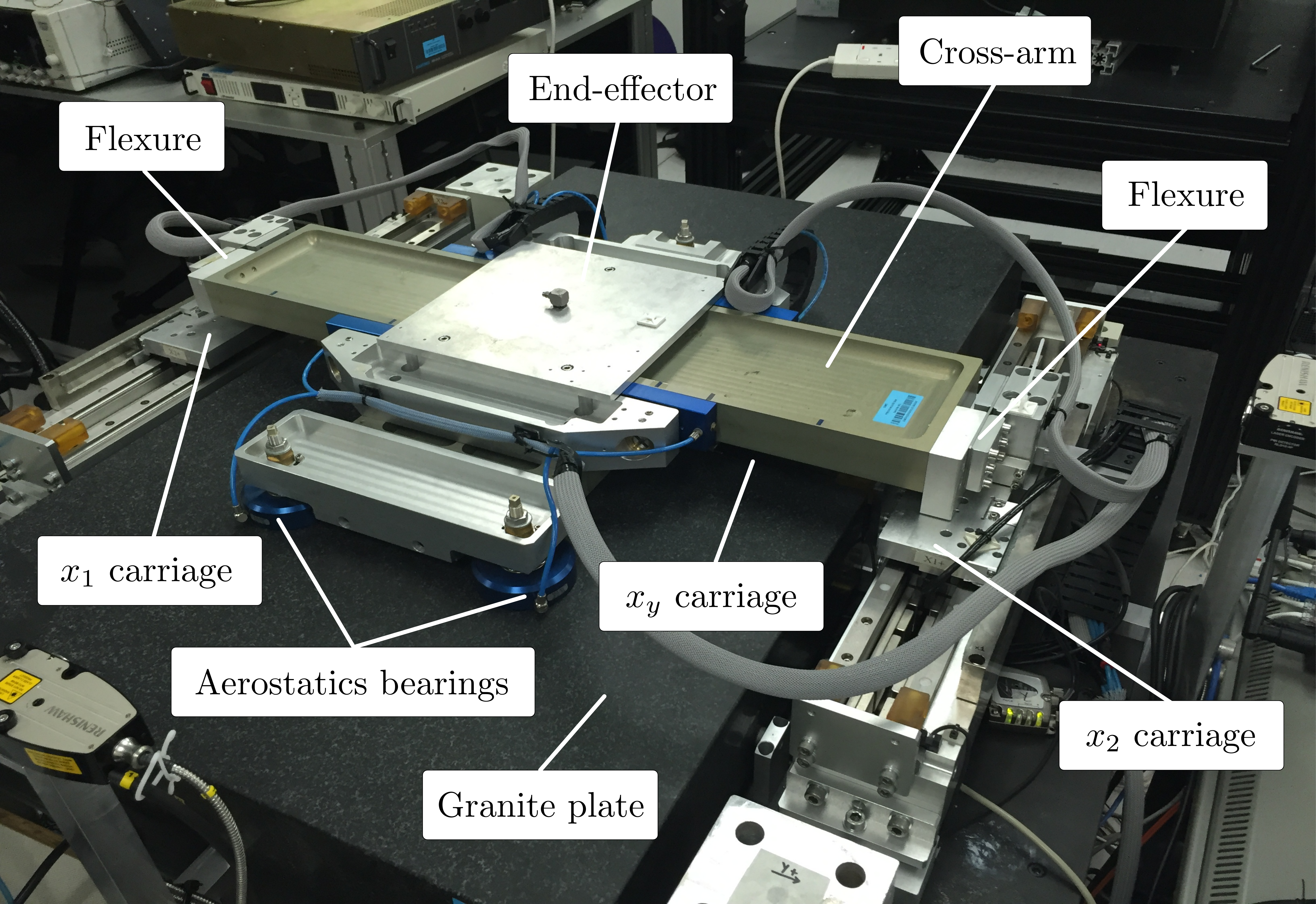}}
	\caption{Air-bearing biaxial gantry system used in this work.}
	\label{fig.gantry}
\end{figure}

This section presents the modeling of the flexure-linked gantry system and also the main issues in contouring motion control tasks. The setup of the air bearing gantry system in this work is shown in Fig.~\ref{fig.gantry}. The gantry system includes three carriages, with two parallel carriages in the $x$-axis and one carriage in the $y$-axis, where each carriage is driven by a linear motor. To ensure the smoothness of the carriages, the platform is supported by several air-bearings. Besides, the entire setup is equipped with a structural bar, a cross-arm, and an end-effector. Specifically, the cross-arm with the end-effector module is linked to the two parallel carriages in the $x$-axis by two flexure joints. These flexure joints are made of flexible stainless steel sheets that allow a small degree of $x$-axis de-synchronization. As a result, the flexible structure provides an additional DOF and protection to the cross-arm, and thus machine-wide damage due to unintended operation can be effectively avoided. Nonetheless, a minor issue resulted from this design is the strong coupling effects, which render the precision contouring control task (especially multi-axis cooperation) more difficult. 

\begin{figure}[t]
	\centerline{\includegraphics[trim = 0cm 0cm 0cm 0cm, width=0.8\columnwidth]{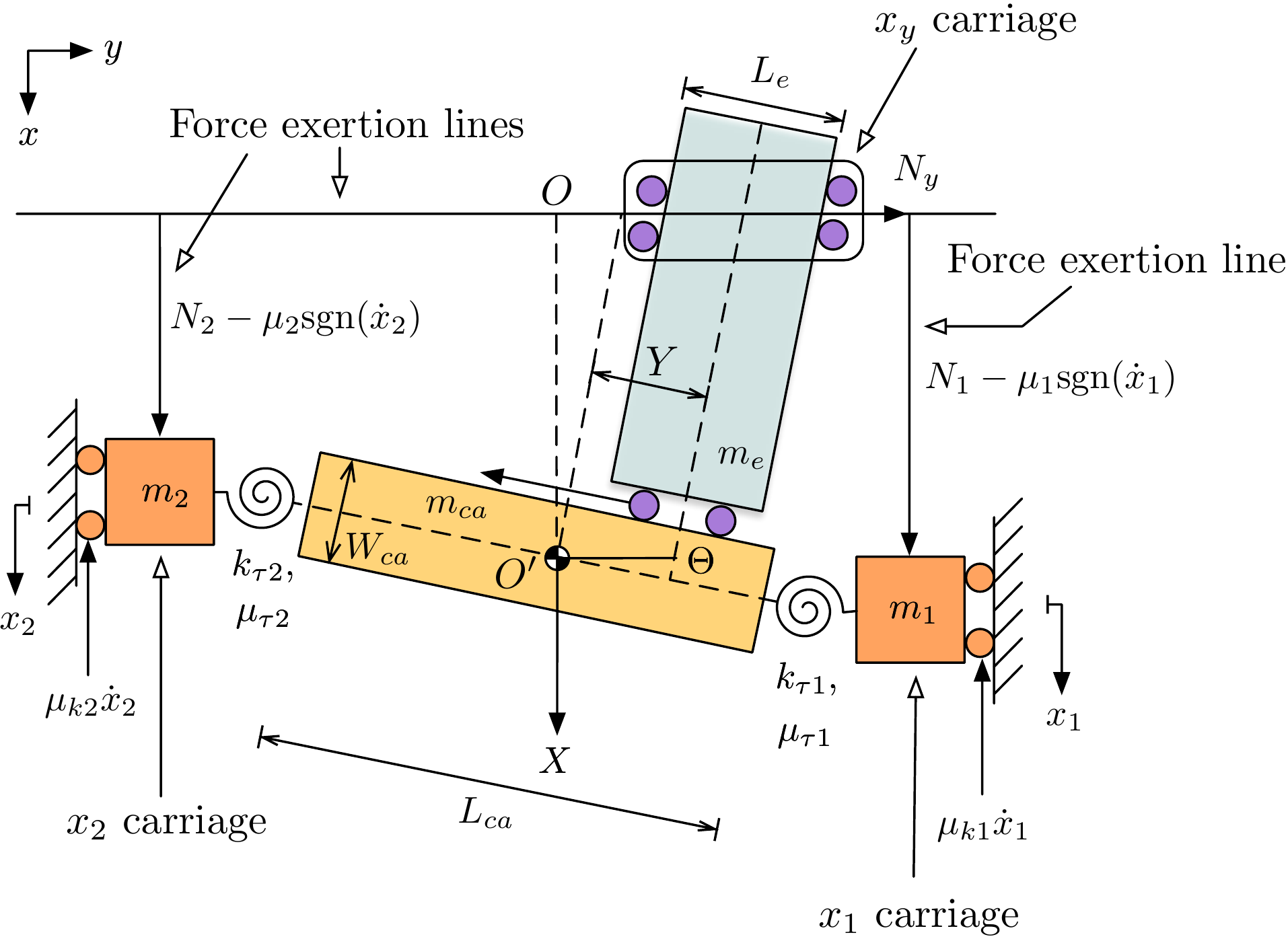}}
	\caption{Schematic diagram of the biaxial gantry system.}
	\label{fig.gantry_model}
\end{figure}

\subsection{Mathematical Model of the Gantry System}
For dynamical modeling of the gantry system, a state vector is defined as $q=\begin{bmatrix}
X & \Theta & Y
\end{bmatrix}^T$, where $X$ and $Y$ denote the position of the cross-arm with respect to the base platform and the position of the end-effector along the cross-arm, respectively; and $\Theta$ represents the orientation of the cross-arm. Also, the schematic diagram of the gantry system is shown in Fig.~\ref{fig.gantry_model} with pertinent symbols listed in Table~\ref{nomenclature}.

\begin{table}\centering	
	\caption{Nomenclature for the Biaxial Gantry System.}	
	\label{nomenclature}       
	\begin{tabular}{lll}		
		\hline\noalign{\smallskip}		
		Symbol & Description & Unit  \\		
		\noalign{\smallskip}\hline\noalign{\smallskip}		
		$m_e$ & Mass of end-effector & kg \\		
		$m_{ca}$ & Mass of cross-arm & kg \\
		$m_1$ & Mass of carriage $x_1$ & kg \\		
		$m_2$ & Mass of carriage $x_2$ & kg  \\		
		$\mu_1$ & Coulomb friction of $m_1$ & N \\
		$\mu_2$ & Coulomb friction of $m_2$ & N \\
		$\mu_y$ & Coulomb friction of $m_e$ & N \\	
		$\mu_{k1}$ & Viscous friction coefficient of $m_1$ & N/(m/s) \\
		$\mu_{k2}$ & Viscous friction coefficient of $m_2$ & N/(m/s) \\	
		$\mu_{kY}$ & Viscous friction coefficient of $m_e$ & N/(m/s) \\		
		$\mu_{\tau 1}$ & Viscous friction of flexure joint 1 & N/(rad/s) \\
		$\mu_{\tau 2}$ & Viscous friction of flexure joint 2 & N/(rad/s) \\		
		$k_{\tau 1}$ & Stiffness of flexure joint 1 & N/rad \\
		$k_{\tau 2}$ & Stiffness of flexure joint 2 & N/rad \\		
		$L_e$ & Length of square shaped end-effector & m \\
		$L_{ca}$ & Length of cross-arm & m \\		
		$W_{ca}$ & Width of cross-arm & m \\
		\noalign{\smallskip}\hline		
	\end{tabular}	
\end{table}

Then, according to Lagrange–Euler formalism, the mathematical model of the gantry system is derived as follows:
\begin{IEEEeqnarray}{rCl}~\label{eq.gantry}
	M(q)\ddot{q}+P(q,\dot{q})\dot{q}+W(q,\dot{q})\dot{q}+Kq+h=N-\gamma
\end{IEEEeqnarray}
where $M,P,W,K\in\mathbb{R}^{3\times 3}$ denote the inertia matrix, the Coriolis and centripetal acceleration matrix, the viscous damping matrix, and the stiffness matrix, respectively. The vectors $h,N,\gamma\in\mathbb{R}^{3}$ refer to the unknown disturbance, the force from control input, and the nonlinear friction (such as Coulomb friction, etc.), respectively. The details of the above matrices and vectors are given in Appendix A.

To analyze the state-space models of the individual parallel carriages, a state vector is defined as $x=\begin{bmatrix}
x_1 & x_2 & x_y
\end{bmatrix}^T$, where $x_1$ and $x_2$ denote the actual positions of the individual parallel carriages
in the $x$-axis; $x_y$ denotes the position of the carriage in the $y$-axis. From Fig.~\ref{fig.gantry_model}, with a small $\Theta$, the following relationships hold

\begin{IEEEeqnarray}{rCl}
	x_1 &=& X+\frac{L_{ca}}{2}{\rm{sin}}(\Theta), x_2 = X-\frac{L_{ca}}{2}{\rm{sin}}(\Theta) \nonumber\\
	X &=& \frac{(x_1+x_2)}{2}, \Theta = \frac{x_1-x_2}{L_{ca}}, Y = x_y{\rm{sec}}(\Theta).\label{eq.modeltrans3}
\end{IEEEeqnarray}

For convenience, the relationship can be written as a coordinate transformation given by
\begin{IEEEeqnarray}{rCl} 
	q = T_px \label{eq.modeltrans}
\end{IEEEeqnarray}
where $T_p=\begin{bmatrix}
1/2 & 1/2 & 0 \\
1/L_{ca} & -1/L_{ca} & 0 \\
0 & 0 & \textup{sec}(\Theta)
\end{bmatrix}$.

Besides, $N=\begin{bmatrix}
N_X & N_\Theta & N_Y
\end{bmatrix}^T$ in~\eqref{eq.gantry} should also be transformed to the forces generated by each individual linear actuator ($\hat{N}=\begin{bmatrix}
N_1 & N_2 & N_y
\end{bmatrix}^T$). The relationships can be easily derived, which are given by
\begin{IEEEeqnarray}{rCl}
	N_1 &=& \frac{N_X-N_Y\textup{sin}(\Theta)}{2}+\frac{N_\Theta}{L_{ca}\textup{cos}(\Theta)} \nonumber\\
	N_2 &=& \frac{N_X-N_Y\textup{sin}(\Theta)}{2}-\frac{N_\Theta}{L_{ca}\textup{cos}(\Theta)} \nonumber\\
	N_y &=& N_Y\textup{cos}(\Theta).\label{eq.forcetrans3}
\end{IEEEeqnarray}

Similarly, \eqref{eq.forcetrans3} can also be written as
\begin{IEEEeqnarray}{rCl} 
	N = T_f\hat{N} \label{eq.forcetrans}
\end{IEEEeqnarray}
where $T_f=\begin{bmatrix}
1 & 1 & \textup{tan}(\Theta) \\
(L_{ca}/2){\rm{cos}}(\Theta) & -(L_{ca}/2){\rm{cos}}(\Theta) & 0 \\
0 & 0 & \textup{sec}(\Theta)
\end{bmatrix}$.

Furthermore, the force generated by each actuator is generally considered as the product of force constant $K_f=\textup{diag}\{K_{f1},K_{f2},K_{fy}\}$ and control input (voltage) $u$. Also, $u=\begin{bmatrix}
u_1 & u_2 & u_y
\end{bmatrix}^T$ denotes the vector consisting of the control input from the three actuators respectively. Therefore, according to \eqref{eq.modeltrans} and \eqref{eq.forcetrans}, \eqref{eq.gantry} can be written as
\begin{IEEEeqnarray}{rCl}~\label{eq.gantry1}
MT_p\ddot{x}+PT_p\dot{x}+WT_p\dot{x}+KT_px+h=T_fK_fu-\gamma. \IEEEeqnarraynumspace
\end{IEEEeqnarray}

\subsection{Main Issues in Contouring Tasks}

To formulate the main problems of the contouring tasks, we define $z=\begin{bmatrix}
x^T & \dot{x}^T
\end{bmatrix}^T=\begin{bmatrix}
x_1 & x_2 & x_y & \dot{x}_1 & \dot{x}_2 & \dot{x}_y
\end{bmatrix}^T$, and \eqref{eq.gantry1} can be written in a regular state-space form
\begin{eqnarray}
\left\{    
\begin{array}{rcl}        
\dot{z} &=& Az+Bu+D \\        
y &=& Cz+w    
\end{array} \right.    \label{ss}
\end{eqnarray}
where $A=\begin{bmatrix}
O_{3\times 3} & I_{3} \\
-(MT_p)^{-1}KT_p & -(MT_p)^{-1}(PT_p+WT_p)
\end{bmatrix}$, $B=\begin{bmatrix}
O_{3\times 3} & ((MT_p)^{-1}T_fK_f)^T
\end{bmatrix}^T$, $D=\begin{bmatrix}
O_{1\times 3} & (h+\gamma)^T
\end{bmatrix}^T$, and $C=\begin{bmatrix}
I_{3} & O_{3\times 3}
\end{bmatrix}$. In this work, $I_n$ represents the identity matrix with a dimension of $n\times n$; and $O_{m\times n}$ represents the all-zero matrix with a dimension of $m\times n$; and the vector $y=\begin{bmatrix}
y_1 & y_2 & y_y
\end{bmatrix}^T$ 

denotes the vector constructed by the actual positions in each axis. Notice that $MT_p$ is always nonsingular. Also, these values are affected by exogenous disturbance $w\in\mathbb{R}^{3}$ (which can certainly include instances of measurement noise); and $w$ can indeed also accommodate the effects of unmatched uncertainties. As discussed above, the matrices $M,P,W$ are state-dependent (related to $q$ and $\dot{q}$), and thus the matrices $A,B,D$ are also related to $z$. Therefore, \eqref{ss} can be regarded as an uncertain system in the following form:
\begin{eqnarray}
\left\{    
\begin{array}{rcl}        
\dot{z} &=& f(z)+g(z)u \\        
y &=& Cz+w    
\end{array}. \right.    \label{ss1}
\end{eqnarray}
In this work, the primary objective is to design a model-free controller for the uncertain multi-axis system, such that a high-precision contouring motion control task can be achieved.

\section{2-DOF Model-Free Controller Design}
\subsection{Synthesis of the Model-Free Controller}
The main target of sliding mode control is to use appropriate signum-type control law to drive and constrain the sliding variable lying within a neighbor of the sliding surface, such that the response and performance of the system can be insensitive to the uncertainty/nonlinearity. Generally, for sliding mode control, the control gains must be designed larger than the upper bound of the uncertainty/nonlinearity. Nevertheless here, such an upper bound is very difficult and impractical to determine. As a possible solution, the control gains can be adaptively tuned without the knowledge of the upper bound of the uncertainty/nonlinearity. 

In this work, the sliding variables are designed as
\begin{IEEEeqnarray}{rCl}~\label{eq.s}
	s = \begin{bmatrix}
		s_1 & s_2 & s_y
	\end{bmatrix}^T = S\begin{bmatrix}
		e & \dot{e}
	\end{bmatrix}^T
\end{IEEEeqnarray}
where $e=\begin{bmatrix}
e_1 & e_2 & e_y
\end{bmatrix}^T$ comprises the tracking error of each axis and $S=\begin{bmatrix}
\textup{diag}\{\lambda_1,\lambda_2,\lambda_y\} & I_3 
\end{bmatrix}$ represents the coefficient matrix of the sliding surface, which should be designed such that they could constitute a Hurwitz polynomial, \emph{i.e.,} $\lambda_1,\lambda_2,\lambda_y>0$. Then, we define the sliding hyperplane $\mathcal{S}$ as
\begin{IEEEeqnarray}{rCl}~\label{eq.hyperplane}
	\mathcal{S}=\left\{   e\in\mathbb{R}^{3} :   s(e)=0\right\}.
\end{IEEEeqnarray}

Then, the control law is designed as
\begin{IEEEeqnarray}{rCl}~\label{eq.u}
	u=-\Gamma\cdot{\rm{sigm}}_a(s)
\end{IEEEeqnarray}
where $\Gamma=\textup{diag}\{\Gamma_1,\Gamma_2,\Gamma_y\}$ consists of time-varying control gains with respect to the sliding variables. Additionally, we have
\begin{IEEEeqnarray}{rCl}~\label{eq.gain}
	\dot{\Gamma}=\bar{\Gamma}\cdot\textup{diag}\{\vert s_1 \vert,\vert s_2 \vert,\vert s_y \vert\}
\end{IEEEeqnarray}
where $\bar{\Gamma}$ is a positive constant.

From \eqref{eq.gain}, it can be seen that the control gains will be larger and larger if the sliding variables are not driven and constrained onto the sliding hyperplane. As a result, the control gains will keep increasing until the sliding dynamics are established, so that the knowledge of the uncertainty/nonlinearity is not needed.

Notice that the function ${\rm{sigm}}_a(s)$ is defined as a sigmoid-like function, where
\begin{IEEEeqnarray}{rCl}~\label{eq.sigm}
     {\rm{sigm}}_a(s)=\frac{1-e^{-as}}{1+e^{-as}}.
\end{IEEEeqnarray}

\begin{remark}
	The sigmoid-like function ${\rm{sigm}}_a(s)$ is a fully continuously differentiable odd function. Differentiate the ${\rm{sigm}}_a(s)$ function with respect to $s$, and it yields
\begin{IEEEeqnarray}{rCl}~\label{eq.sigm_dot}
\frac{d\:{\rm{sigm}}_a(s)}{ds}=2a\:{\rm{sigm}}_a(s)(1-{\rm{sigm}}_a(s))
\end{IEEEeqnarray}
which exists for all $s\in\mathbb{R}$, and the parameter $a$ can be used to tune the switching rate in practical applications. Note that ${\rm{sigm}}_a(s)$ (rather than ${\rm{sgn}}(s)$) functions as the switching operator in the control law due to its smoothness and differentiability. 

In conventional sliding mode control approaches where the non-continuous ${\rm{sgn}}(s)$ function is used, the switching rates of the sliding mode control could likely (in certain unfortunate situations) turn out to be close to the natural frequencies (resonances) of the system, and thus the chattering issues could be resultingly severe. In our work, with the alternative utilization of the smoother sigmoid-like function ${\rm{sigm}}_a(s)$ (which is a fully continuously differentiable function), the afore-mentioned undesirable phenomenon as typically appears in conventional sliding mode control is essentially effectively avoided.
\end{remark}

Although the adaptive sliding mode control is effective, there are still some drawbacks from the practical point of view. As one of the general properties of sliding mode control, the ideal sliding motion can never be realized. As a result, the control gains are always increasing, and thus the overestimation of the control gains is inevitable. Moreover, larger control gains lead to higher chattering in sliding mode control, and this adaptation law will consequently bring unnecessary chattering into the system. Meanwhile, it can also be seen in \eqref{eq.gain} that the control gains will increase with a small rate when the sliding variables are closer to the sliding hyperplane. Hence, the precision of the tracking performance is not easy to be guaranteed. Besides, this method can only deal with matched uncertainties while unmatched uncertainties are still required to be suppressed in this work. Here, an iterative learning law is used to compensate for the repetitive uncertainties (both matched and unmatched ones). The iterative learning law in the $i$-th iteration with a learning rate $l$ is designed as follows:
\begin{IEEEeqnarray}{rCl}
y_{r,i+1} &=& r+w_{i+1} \label{eq.ILC_law1} \\
w_{i+1}   &=& w_i+le_{i+1} \label{eq.ILC_law2}
\end{IEEEeqnarray}
where $r\in\mathbb{R}^3$ is the reference signal vector of the three axes, $w_i\in\mathbb{R}^3$ is the iterative variable vector during the consecutive iterations. \eqref{eq.ILC_law1} and \eqref{eq.ILC_law2} show that the iterative learning control input will be added to the reference signal. Generally, a low-pass filter will be used in the iterative learning law, and thus the original reference signal is not allowed to be changed by the filter. With the iterative learning control and a low-pass filter, the reference signal after learning is not distorted with extra undesirable magnitude change and phase lag.

\subsection{Overview of the Control Structure}

\begin{figure}[t]
	\centerline{\includegraphics[trim = 0cm 0cm 0cm 0cm, width=\columnwidth]{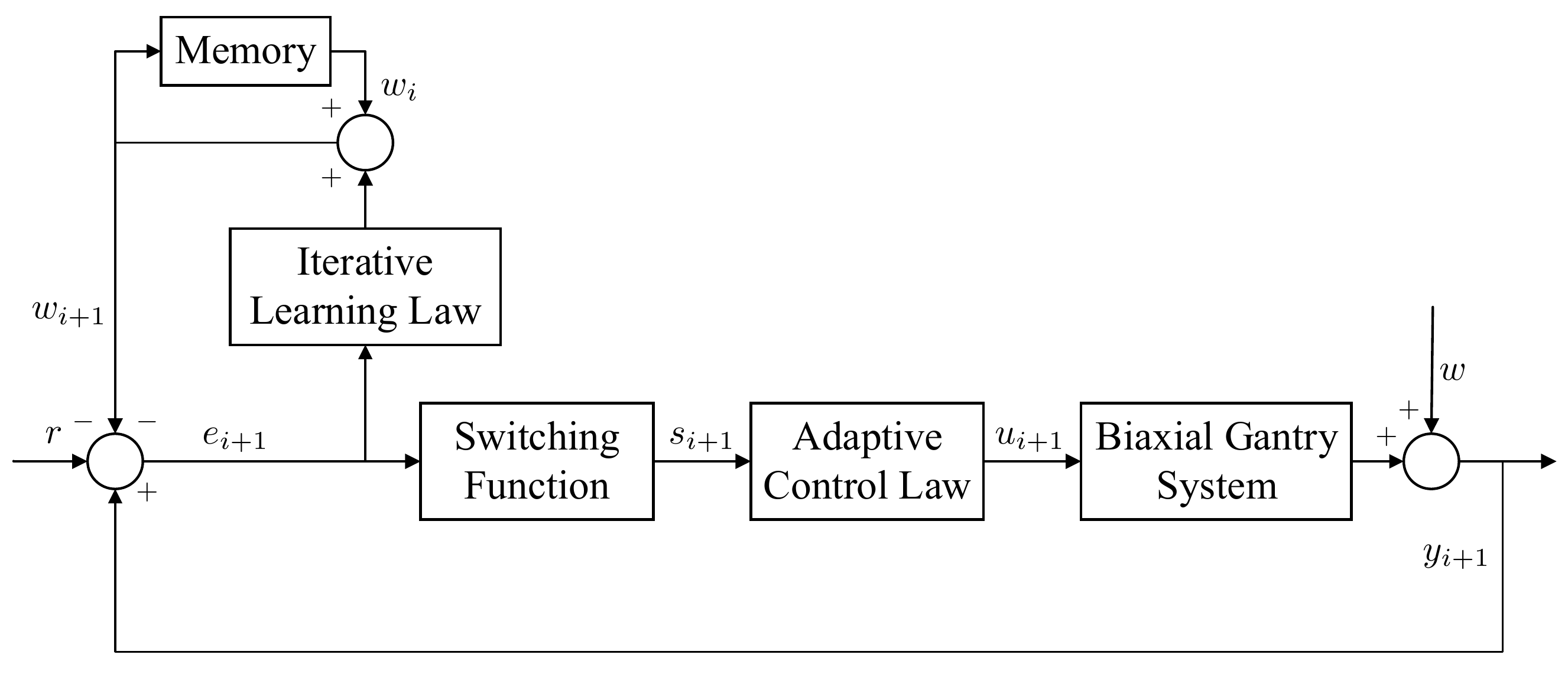}}
	\caption{Overall control architecture with the proposed methodology.}
	\label{fig.diagram}
\end{figure}

The overall structure of the proposed control scheme is shown in Fig.~\ref{fig.diagram}. For each carriage, all of the effects from other carriages and actuators are regarded as uncertainties; and then, each carriage tracks its own reference with high precision to realize high contouring accuracy. The key idea of the proposed method is to use varying-gain sliding mode control to suppress the matched uncertainties (including nonlinearity, coupling, and disturbance) and use iterative learning law to compensate for the repetitive matched uncertain dynamics as well as the unmatched disturbance. As a result, the adaptive control gain $\Gamma$ is not needed to keep increasing for better performance and the problem of control gain overestimation is solved in the iteration domain. In other words, the control gains at the same level can make the sliding variables closer to the sliding hyperplane $\mathcal{S}$ in the iteration domain, which can also be seen in Section IV. Then, these properties are illustrated theoretically below.

Recall that the system in \eqref{ss1} is described as follows
\begin{equation}
\left\{
\begin{array}{rcl}
\dot{z} &=& f(z)+g(z)u \\
y &=& Cz+w
\end{array}. \right. \label{ss2}
\end{equation}
Substitute \eqref{eq.ILC_law1} and \eqref{eq.ILC_law2} into \eqref{ss2}, we have
\begin{IEEEeqnarray}{rCl}
e_i &=& y_i-y_{r,i} \nonumber \\
&=& Cz_i+w-r-w_i \\
\dot{e}_i &=& C\dot{z}_i+\dot{w}-\dot{r}-\dot{w}_i \nonumber \\ 
&=& C(f(z_i)+g(z_i)u_i)-\dot{r}+\dot{w}-\dot{w}_i.
\end{IEEEeqnarray}
Consequently, from \eqref{eq.s}, the sliding variable in the $i$-th iteration is given by
\begin{IEEEeqnarray}{rcl} 
s_i &=& \textup{diag}\{\lambda_1,\lambda_2,\lambda_y\} e_i+\dot{e}_i \nonumber\\
&=& \textup{diag}\{\lambda_1,\lambda_2,\lambda_y\} (Cz_i+w-r-w_i)+C\dot{z}_i-\dot{r}+\dot{w}-\dot{w}_i \nonumber\\
&=& F(z_i)+G(z_i)u_i+\phi-\phi_i  \nonumber\\
&=& v_i+\varpi_i \label{s_dot}
\end{IEEEeqnarray}
where $F(z_i)\in\mathbb{R}^3$ consists of the unknown functions of the three axes with respect to $t$ and $z_i$, \emph{i.e.,} $F(z_i)=\begin{bmatrix}
F_1(z_i) & F_2(z_i) & F_y(z_i)
\end{bmatrix}^T$. And similarly, $G(z_i)u_i\in\mathbb{R}^3$ can be also written in  uncertain dynamics form, \emph{i.e.,} $G(z_i)u_i=\begin{bmatrix}
G_1(z_i,u_i)u_{1,i} & G_2(z_i,u_i)u_{2,i} & G_y(z_i,u_i)u_{y,i}
\end{bmatrix}^T$. For convenience, $v_i=F(z_i)+G(z_i)u_i$ and $\varpi_i=\phi-\phi_i$, $\phi=\lambda w+\dot{w}$, $\phi_i=\lambda w_i+\dot{w}_i$ are defined. 

\begin{remark}  \label{bound}
Define that $(\star)_{n,i}$ denotes the variable $\star$ of the $n=1,2,y$ carriage in the $i$-th iteration. $F(z_i)$ and $G(z_i,u_i)$ are regarded as unknown matrices. Note that in practical reference tracking problems, the length of the reference signal is always finite, \emph{i.e.,} $t\in\left[0,T\right]$. Therefore, there must exist element-wise upper bounds $F_{M,n,i}$ of the unknown functions $F_n(z_i)$ within the reference signal length, \emph{i.e.,} $\vert F_n(z_i)\vert\leq F_{M,n,i}$ for all $t\in\left[0,T\right]$. Similarly, $G(z_i,u_i)$ has $0<G_{m,n,i}<G_n(z_i,u_i)\leq G_{M,n,i}$ for all $t\in\left[0,T\right]$. Hence, there must exist gains $\Gamma_n^*$ that satisfy $\Gamma_{n,i}^*>\frac{F_{M,n,i}}{G_{m,n,i}}\geq\vert\frac{F_{n}(z_i)}{G_{n}(z_i,u_i)}\vert$.
\end{remark}

\subsection{Convergence and Stability Analysis}
To analyze the effectiveness of the proposed method, the convergence in the iteration domain and the stability (sliding motion is established in finite time) in the time domain are proved in this section. At first, the stability in each iteration is analyzed. Notice that the iterative learning law is only added to the reference signal. Therefore, for each iteration, the task is actually a tracking problem with $r'=r+w_i$ under the control law \eqref{eq.u} and the adaptation law \eqref{eq.gain}.  

\begin{theorem} \label{theorem_sta}
	For an uncertain system $\dot{z}=\bar{f}(z)+\bar{g}(z,u)u$ with control law \eqref{eq.u} and adaptation law \eqref{eq.gain}, there exists a finite time $t_F>0$ so that the sliding motion is established for all $t\geq t_F$.  
\end{theorem}

\noindent{\textbf{Proof of Theorem \ref{theorem_sta}:}} First, we have
\begin{equation}
\begin{aligned}
\dot{s} &= \frac{\partial s}{\partial t}+\frac{\partial s}{\partial z}\dot{z}\\
&= \underbrace{\frac{\partial s}{\partial t}+\frac{\partial s}{\partial z} \bar f(z)}+\underbrace{\frac{\partial s}{\partial z} \bar g(z,u)}u. \\
& \quad \quad \quad  \bar{F}(z) \qquad \quad \:\: \bar{G}(z,u)
\label{sigma-dynamics1}
\end{aligned}
\end{equation}
Then the Lyapunov function is designed as
\begin{IEEEeqnarray}{rCl}
	J &=& \sum_{n=1,2,y}J_n \nonumber\\
	  &=& \sum_{n=1,2,y}\frac{1}{2}s_n^2+\sum_{n=1,2,y}\frac{1}{2\delta}(\Gamma_{n}-\Gamma_n^*)^2 
\end{IEEEeqnarray}
where $\delta$ is a positive constant. By differentiation, it follows that
\begin{equation}
\begin{aligned}
\dot{J}_n =& s_n\dot{s}_n+\delta^{-1}(\Gamma_n-\Gamma_n^*)\dot{\Gamma}_n \\
        =& s_n(\bar{F}_n+\bar{G}_nu_n)+\delta^{-1}\bar{\Gamma}_n(\Gamma_n-\Gamma_n^*)\vert s_n\vert \\
        \leq& \vert s_n\vert \bar{F}_{M,n}-\bar{G}_{m,n} \Gamma_n \vert s_n\vert+\delta^{-1}\bar{\Gamma}_n(\Gamma_n-\Gamma_n^*)\vert s_n\vert \\
        =& \vert s_n\vert (\bar{F}_{M,n}-\bar{G}_{m,n}\Gamma_n^*)+(\Gamma_n-\Gamma_n^*)(\delta^{-1}\bar{\Gamma}_n-\bar{G}_{m,n})\vert s_n\vert.
\end{aligned}
\end{equation}
Define $\alpha=-\bar{F}_{M,n}+\bar{G}_{m,n}\Gamma_n^*$ and $\beta=-(\delta^{-1}\bar{\Gamma}_n+\bar{G}_{m,n})\vert s_n\vert$. Then, from \eqref{eq.u}, \eqref{eq.gain}, and Remark \ref{bound}, we can find that $\alpha>0$ and $\Gamma_n-\Gamma_n^*<0$ for all $t>0$. Also, it is easy to find a positive constant $\delta$ such that $\beta>0$. Therefore, it yields
\begin{IEEEeqnarray}{rcl}
\dot{J}_n &\leq& -\alpha\vert s_n\vert-\beta\vert\Gamma_n-\Gamma_n^*\vert \nonumber\\
          &\leq& -\text{min}\{\sqrt{2}\alpha,\sqrt{2\delta}\beta\}\bigg(\frac{\vert s_n\vert}{\sqrt{2}}+\frac{\vert\Gamma_n-\Gamma_n^*\vert}{\sqrt{2\delta}}\bigg) \nonumber\\
          &\leq& -\gamma \sqrt{J_n}
\end{IEEEeqnarray}
where $\gamma=\text{min}\{\sqrt{2}\alpha,\sqrt{2\delta}\beta\}$. Thus, $\dot{J}_n$ is non-positive so that the establishment of the sliding motion is proved and the finite reaching time $t_F$ can be estimated as $t_F\leq{2\sqrt{J_n(0)}}/{\gamma}$. Therefore, $\dot{J}$ (the sum of $\dot{J}_n$) is consequently non-positive and thus the stability is proved. \hfill{\qedsymbol}

As a crucially important property, the convergence within the iterations is proved in Theorem \ref{theorem_con}.

\begin{theorem} \label{theorem_con}
	 The system (\ref{ss2}) with control law (\ref{eq.u}), (\ref{eq.gain}), and learning rule (\ref{eq.ILC_law1}) is convergent as
\begin{eqnarray} 
&(i) \quad &\lim_{i\rightarrow\infty}V_i(t)=V(t)  \\
&(ii) \quad & \lim_{i\rightarrow\infty}s_i(t)=0  \quad {\rm{for }}\:\forall t\in[0,T] 
\end{eqnarray}
where $V_i(t)$ is defined as a Lyapunov-like learning performance function and the time index $t$ is omitted for convenience.
\begin{equation}
V_i=V_i^{\langle 1\rangle}+V_i^{\langle 2\rangle}+V_i^{\langle 3\rangle} \label{V}
\end{equation}
with
\begin{IEEEeqnarray}{rcl} 
V_i^{\langle 1\rangle} &=& \sum_{n=1,2,y}V_{n,i}^{\langle 1\rangle} = \sum_{n=1,2,y}\int_{0}^{T}\frac{1}{2}s_{n,i}(\tau)^Ts_{n,i}(\tau)\:d\tau  \label{V1}\\
V_i^{\langle 2\rangle} &=& \sum_{n=1,2,y}V_{n,i}^{\langle 2\rangle} = \sum_{n=1,2,y}\int_{0}^{T}\frac{1}{2}l^{-1}\varpi_{n,i}(\tau)^T\varpi_{n,i}(\tau)\:d\tau \nonumber \\ \label{V2}\\
V_i^{\langle 3\rangle} &=& \sum_{n=1,2,y}V_{n,i}^{\langle 3\rangle} = \sum_{n=1,2,y}\frac{1}{2\delta}(\Gamma_{n,i}-\Gamma_n^*)^2. \label{V3}
\end{IEEEeqnarray}
\end{theorem}
\noindent{\textbf{Proof of Theorem \ref{theorem_con}:}}  Consider the function (\ref{V}), the difference of $V_i^{\langle 1\rangle}$, $V_i^{\langle 2\rangle}$, $V_i^{\langle 3\rangle}$ between two consecutive iterations $(i+1)$ and $(i)$ is derived respectively. Notice that the defined performance function $V$ is actually the sum of the variables of the three axes. Hence, we only analyze one of the axes in the following equations. For convenience, the lower subscript $n$ is omitted in the following analysis.

Consider (\ref{V1}) with (\ref{s_dot}), we have
\begin{equation} 
\begin{aligned}
\triangle V_i^{\langle 1\rangle}
=& \int_{0}^{T}s_{i+1}(\tau)(v_{i+1}(\tau)+\varpi_{i+1}(\tau))\:d\tau \\ -&\int_{0}^{T}\frac{1}{2}s_i(\tau)^Ts_i(\tau)\:d\tau \\     
=& \int_{0}^{T}s_{i+1}(\tau)(F+Gu_{i+1}(\tau))\:d\tau \\ +& \int_{0}^{T}s_{i+1}(\tau)\varpi_{i+1}(\tau)\:d\tau-\int_{0}^{T}\frac{1}{2}s_i(\tau)^Ts_i(\tau)\:d\tau.
\end{aligned}
\label{V1_1}      
\end{equation}
The difference of $V_i^{\langle 2\rangle}$ between two consecutive iterations $(i+1)$ and $(i)$ is given by
\begin{equation} 
\begin{aligned} 
\triangle V_i^{\langle 2\rangle} =& \int_{0}^{T}\frac{1}{2}l^{-1}\varpi_{i+1}(\tau)^T\varpi_{i+1}(\tau)\:d\tau\\-&\int_{0}^{T}\frac{1}{2}l^{-1}\varpi_i(\tau)^T\varpi_i(\tau)\:d\tau \\
=& \frac{1}{2l}\int_{0}^{T}(\phi_i(\tau)-\phi_{i+1}(\tau))^T\big(2(\phi(\tau)-\phi_{i+1}(\tau)) \\+&(\phi_{i+1}(\tau)-\phi_i(\tau))\big)\:d\tau \\
=& l^{-1}\int_{0}^{T}(\phi_i(\tau)-\phi_{i+1}(\tau))^T(\phi(\tau)-\phi_{i+1}(\tau))\:d\tau \\ +&\frac{1}{2l}\int_{0}^{T}(\phi_i(\tau)-\phi_{i+1}(\tau))^T(\phi_{i+1}(\tau)-\phi_i(\tau))\:d\tau.
\end{aligned}
\label{V22}    
\end{equation}

Consider \eqref{V22} with \eqref{eq.ILC_law1}, $\triangle V_i^{\langle 2\rangle}$ can be written in the following form:
\begin{equation} 
\begin{aligned} 
\triangle V_i^{\langle 2\rangle}=& l^{-1}\int_{0}^{T}(\lambda w_i(\tau)+\dot{w}_i(\tau)-\lambda w_{i+1}(\tau)-\dot{w}_{i+1}(\tau))^T\\&(\phi(\tau)-\phi_{i+1}(\tau))\:d\tau \\ -&\frac{1}{2l}\int_{0}^{T}(\phi_{i+1}(\tau)-\phi_i(\tau))^T(\phi_{i+1}(\tau)-\phi_i(\tau))\:d\tau \\
=& -\int_{0}^{T}s_{i+1}(\tau)^T\varpi_{i+1}(\tau)\:d\tau \\-&\frac{l}{2}\int_{0}^{T}s_{i+1}(\tau)^Ts_{i+1}(\tau)\:d\tau.
\label{V2_1}
\end{aligned}       
\end{equation}
The difference of $V_i^{\langle 3\rangle}$ between the $(i+1)$-th and $(i)$-th iterations can be described as
\begin{equation} 
\begin{aligned} 
\triangle V_i^{\langle 3\rangle} =& \frac{1}{2\delta}(\Gamma_{i+1}-\Gamma^*)^2-\frac{1}{2\delta}(\Gamma_i-\Gamma^*)^2 \\
=& \frac{1}{2\delta}\int_{0}^{T}(\Gamma_{i+1}(\tau)-\Gamma^*)\dot{\Gamma}_{i+1}(\tau)\:d\tau-\frac{1}{2\delta}(\Gamma_i-\Gamma^*)^2.
\end{aligned}   
\label{KK*}  
\end{equation}
From~\eqref{eq.gain}, we have
\begin{equation} 
\triangle V_i^{\langle 3\rangle}=\int_{0}^{T}(\Gamma_{i+1}(\tau)-\Gamma^*)\bar{\Gamma}\vert s_{i+1}(\tau)\vert\:d\tau-\frac{1}{2\delta}(\Gamma_i-\Gamma^*)^2.
\end{equation}
Adding the equations (\ref{V1_1}), (\ref{V2_1}), and (\ref{KK*}), the difference of the learning performance function between the $(i+1)$-th and $i$-th iterations can be written as follows:
\begin{IEEEeqnarray}{rcl}
\triangle V_i &=& \triangle V_i^{\langle 1\rangle}+\triangle V_i^{\langle 2\rangle}+\triangle V_i^{\langle 3\rangle} \nonumber\\
&=& \int_{0}^{T}s_{i+1}(\tau)(F+Gu_{i+1}(\tau))\:d\tau \nonumber\\ &\:&+\int_{0}^{T}(\Gamma_{i+1}(\tau)-\Gamma^*)\bar{\Gamma}\vert s_{i+1}(\tau)\vert\:d\tau \nonumber\\ &\:&-\frac{1}{2\delta}(\Gamma_i-\Gamma^*)^2-\frac{l}{2}\int_{0}^{T}s_{i+1}(\tau)^Ts_{i+1}(\tau)\:d\tau \nonumber\\&\:&-\int_{0}^{T}\frac{1}{2}s_i(\tau)^Ts_i(\tau)\:d\tau \nonumber\\  
&\leq& \int_{0}^{T}s_{i+1}(\tau)(F+Gu_{i+1}(\tau))\:d\tau\nonumber\\&\:&+\int_{0}^{T}(\Gamma_{i+1}(\tau)-\Gamma^*)\bar{\Gamma}\vert s_{i+1}(\tau)\vert\:d\tau.  
\end{IEEEeqnarray}
Substitute the control law (\ref{eq.u}), it yields
\begin{IEEEeqnarray} {rcl}
\triangle V_i &\leq&  \int_{0}^{T}\bigg(s_{i+1}(\tau)\big(F-G\Gamma_{i+1}(\tau)\text{sigm}_a(s_{i+1}(\tau))\big)\nonumber\\&\:&+(\Gamma_{i+1}(\tau)-\Gamma^*)\bar{\Gamma}\vert s_{i+1}(\tau)\vert\bigg)\:d\tau \nonumber\\
&=& \int_{0}^{T}\dot{J}_{i+1}(\tau)\:d\tau  . 
\end{IEEEeqnarray}
In Theorem \ref{theorem_sta}, $\dot{J}_{i+1}\leq0$ has been proved, and thus it yields
\begin{equation}
\triangle V_i \leq \int_{0}^{T}\dot{J}_{i+1}(\tau)\:d\tau\leq0.
\end{equation}
Similarly, $\triangle V_{n,i}$ for $n=1,2,y$ are all non-positive and thus $\triangle V_i$ (sum of $\triangle V_{n,i}$) is also non-positive. Therefore, the convergence in the iteration domain is proved, and also, the sliding variables $s_i$ will reach the sliding hyperplane $\mathcal{S}$ in the iteration domain. \hfill{\qedsymbol}

\section{Real-time Experiment}

\subsection{Experiment Setup}
In this section, the proposed global iterative sliding mode control method is applied to the industrial biaxial gantry stage shown in Fig. \ref{fig.gantry}; where the three moving carriages are driven by three Akribis AUM3-S4 linear motors respectively and each linear motor is connected to a Copley Accelus driver for power; also, a dSPACE DS1105 is used to implement the real-time control with a sampling frequency of 20 kHz. Besides, linear encoders with a resolution of 0.5 $\mu$m  are used for position measurement. 

In this work, the system parameters and the knowledge of uncertainties are unknown. The parameter of the adaptation law \eqref{eq.gain} should be set as a relatively large value, and here, we set $\bar{\Gamma}=1000$ and the initial value of control gains $\Gamma(0)=\textup{diag}\{1,1,1\}$ to make the sigmoid function close to a signum function, the parameter $a$ in control law \eqref{eq.u} is set as $a=10$; the parameters of sliding hyperplane are set as $\lambda_1=\lambda_2=\lambda_y=6$; the learning rate in \eqref{eq.ILC_law2} is set as $l=1$. Notice that the parameters are kept invariant in all experiments (especially all iterations in one group of experiments).  

For the contouring tasks, the parameterized functions (the reference signal of X-axis $x_d$ and Y-axis $y_d$ with respect to time $t$) of the desired contour are given at first, and then the carriages $x_1$ and $x_2$ track the reference signal of $x_d$ and the carriage $x_y$ tracks the reference signal of $y_d$.  
\subsection{Performance Indexes}

In this work, several indexes will be used to evaluate the quality of the performance. Firstly, the contouring error is defined as $e_c = {f(x,y)}/{\sqrt{f_x^2+f_y^2}}
$, where $f(x,y)$ denotes substituting the actual position $(x,y)$ into the desired contour function $f(x_d,y_d)=0$, with $f_x={\partial f(x,y)}/{\partial x}$ and $f_y={\partial f(x,y)}/{\partial y}$. To illustrate the convergence ability of the proposed method, the root-mean-square value of the contouring error and the sliding variables of the three carriages are defined as $\Vert e_c\Vert_{\textup{rms}}=\sqrt{\frac{1}{T}\int_{0}^{T}\vert e_c\vert^2\:dt}$ (RMSE), $\Vert s_1\Vert_{\textup{rms}}=\sqrt{\frac{1}{T}\int_{0}^{T}\vert s_1\vert^2\:dt}$ (RMSSV-$s_1$), $\Vert s_2\Vert_{\textup{rms}}=\sqrt{\frac{1}{T}\int_{0}^{T}\vert s_2\vert^2\:dt}$ (RMSSV-$s_2$), and $\Vert s_y\Vert_{\textup{rms}}=\sqrt{\frac{1}{T}\int_{0}^{T}\vert s_y\vert^2\:dt}$ (RMSSV-$s_y$), respectively. Besides, MaxAE and MaxASV represent the maximum absolute value of error and the maximum absolute value of sliding variable, respectively.

\begin{figure}[t]
	\centerline{\includegraphics[trim = 0.2cm 0cm 0.4cm 1cm, width=6.5cm]{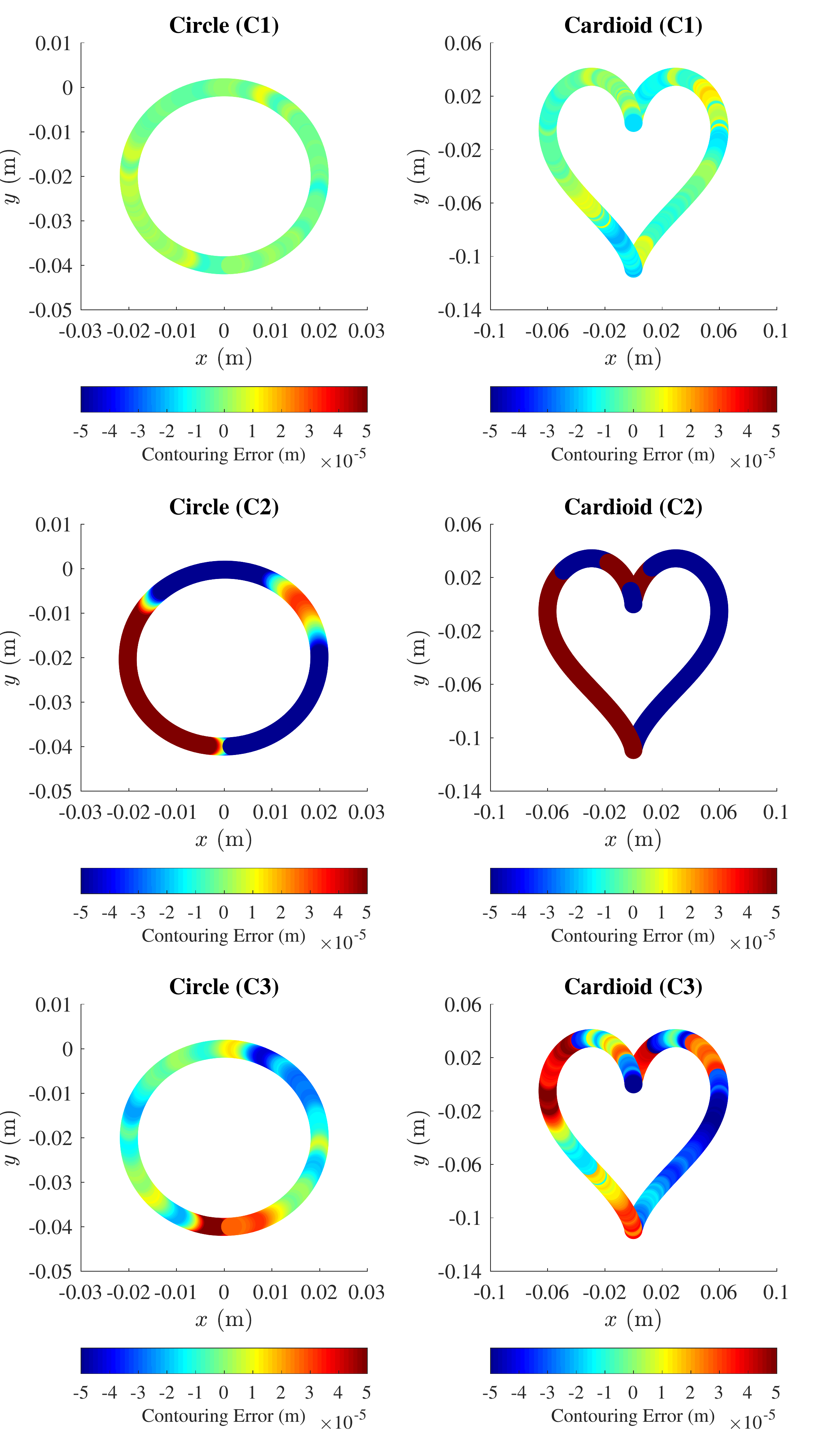}}
	\caption{\textbf{Top-Left.} The contouring error of Task 1 under Controller 1. \textbf{Top-Right.} The contouring error of Task 2 under Controller 1. \textbf{Center-left.} The contouring error of Task 1 under Controller 2. \textbf{Center-Right.} The contouring error of Task 2 under Controller 2. \textbf{Bottom-Left.} The contouring error of Task 1 under Controller 3. \textbf{Bottom-Right.} The contouring error of Task 2 under Controller 3.}
	\label{fig.contour}
\end{figure}

\subsection{Experimental Results}

To demonstrate the contouring performance and the effectiveness of the proposed method, the tasks with two different contours are conducted with the same controller parameters, and 6 iterations are executed in both tasks. Moreover, the contour functions $f(x_d,y_d)$ are shown in Appendix B. 

\begin{figure}[t]
	\centerline{\includegraphics[trim = 0cm 0cm 0cm 1cm, width=0.9\columnwidth]{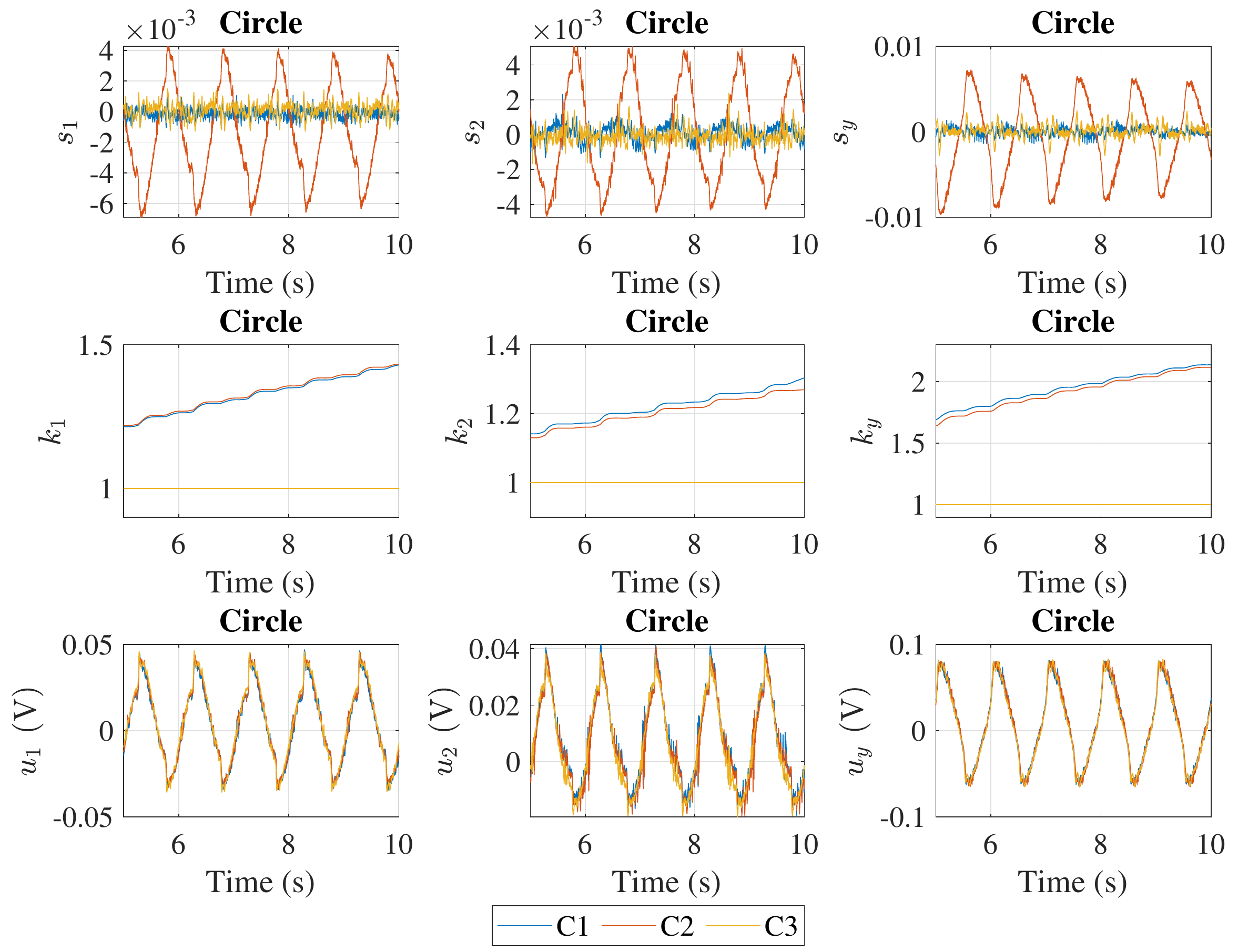}}
	\caption{The sliding variables, control gains, and control inputs of each axis respectively in Task 1 compared between Controller 1, Controller 2, and Controller 3.}
	\label{fig.circle_sku}
\end{figure}

\begin{figure}[t]
	\centerline{\includegraphics[trim = 0cm 0cm 0cm 1cm, width=0.9\columnwidth]{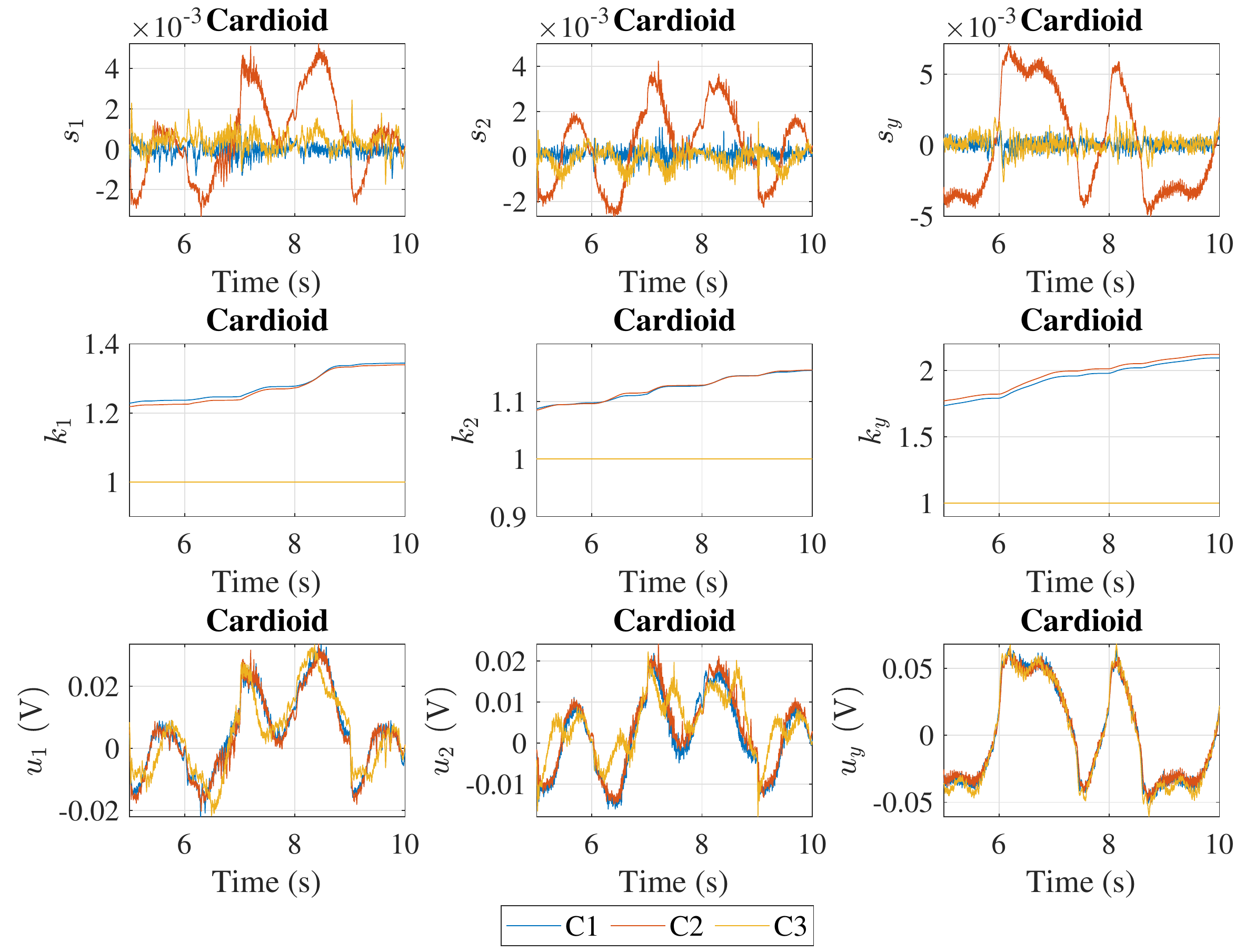}}
	\caption{The sliding variables, control gains, and control inputs of each axis respectively in Task 2 compared between Controller 1, Controller 2, and Controller 3.}
	\label{fig.cardioid_sku}
\end{figure}

\begin{figure}[t]
	\centerline{\includegraphics[trim = 1.0cm 1cm 1.4cm 1.1cm, width=0.73\columnwidth]{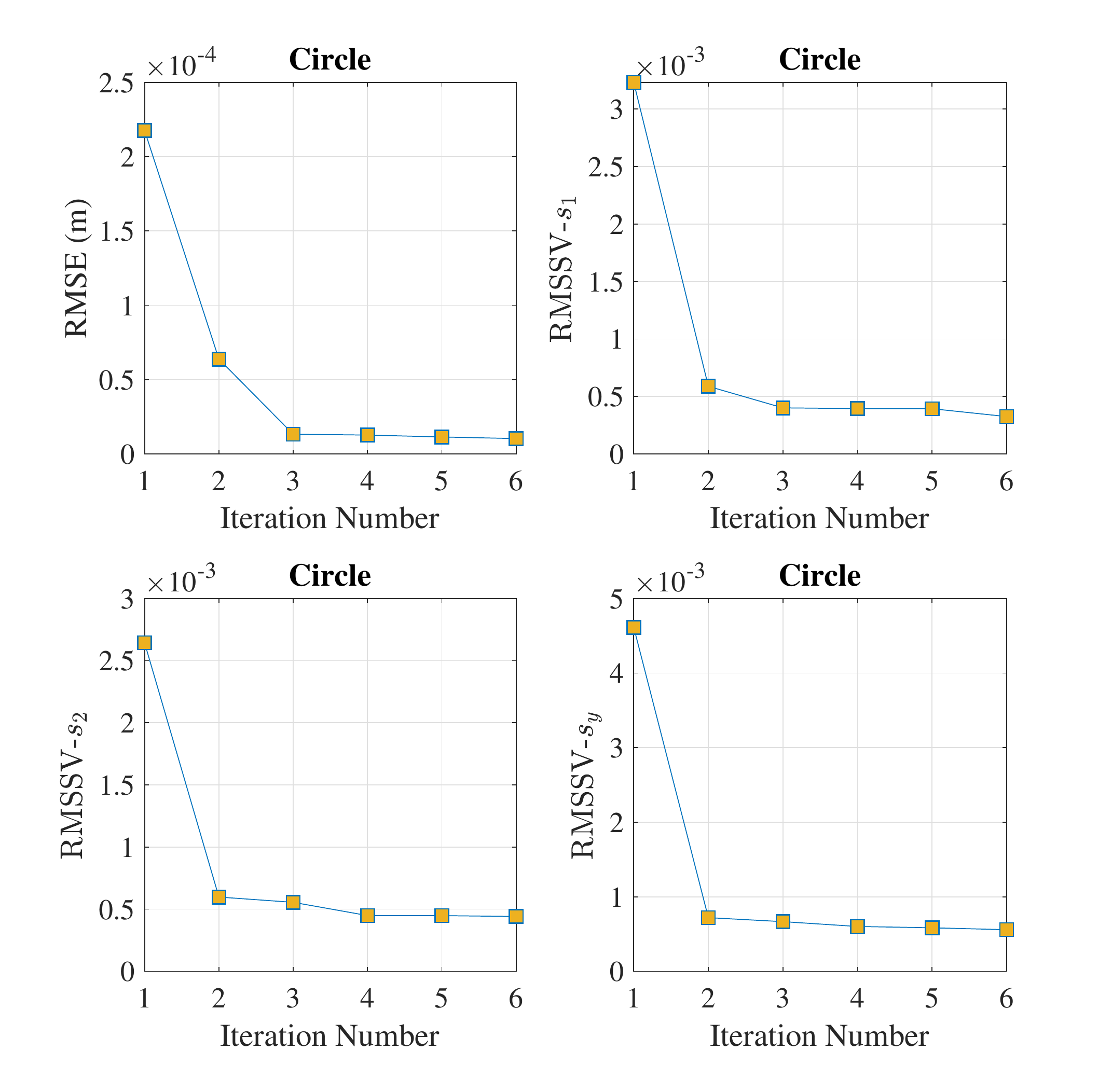}}
	\caption{\textbf{Convergence of contouring error and sliding variables in the first 6 consecutive iterations (Task 1).} \textbf{Top-Left.} The RMS values of the contouring error. \textbf{Top-Right.} The RMS values of the sliding variable $s_1$. \textbf{Bottom-Left.} The RMS values of the sliding variable $s_2$. \textbf{Bottom-Right.} The RMS values of the sliding variable $s_y$.}
	\label{fig.converge_circle}
\end{figure}

\begin{figure}[t]
	\centerline{\includegraphics[trim = 1.0cm 1cm 1.4cm 1.1cm, width=0.73\columnwidth]{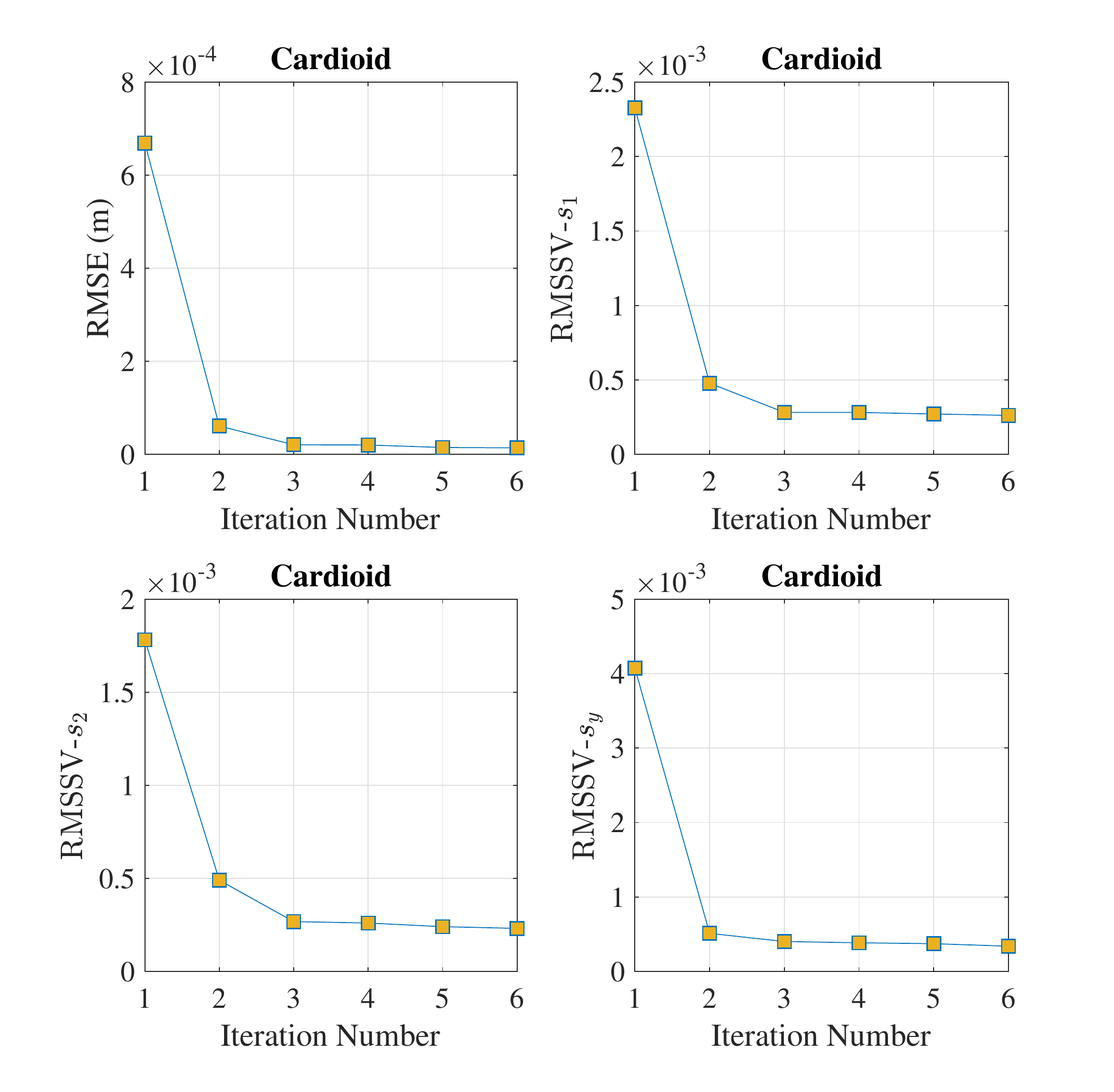}}
	\caption{\textbf{Convergence of contouring error and sliding variables in the first 6 consecutive iterations (Task 2).} \textbf{Top-Left.} The RMS values of the contouring error. \textbf{Top-Right.} The RMS values of the sliding variable $s_1$. \textbf{Bottom-Left.} The RMS values of the sliding variable $s_2$. \textbf{Bottom-Right.} The RMS values of the sliding variable $s_y$.}
	\label{fig.converge_card}
\end{figure}

Task 1 (T1): Circle contour 
\begin{eqnarray}
\left\{    
\begin{array}{rcl}        
x_d(t) &=& 0.02\textup{sin}(2\pi t) \\        
y_d(t) &=& 0.02\textup{cos}(2\pi t)-0.02  .  
\end{array} \right.    \nonumber
\end{eqnarray}

Task 2 (T2): Cardioid contour
\begin{eqnarray}
\left\{    
\begin{array}{rcl}        
x_d(t) &=& 0.06\textup{sin}^3(0.5\pi t) \\        
y_d(t) &=& 0.065\textup{cos}(0.5\pi t)-0.025\textup{cos}(\pi t) \\&\:&- 0.01\textup{cos}(1.5\pi t) - 0.005\textup{cos}(2\pi t) -0.025.
\end{array} \right.  \nonumber 
\end{eqnarray}

To show the effectiveness and superiority of the proposed approach, this paper compares the global iterative sliding mode controller with two other methods. One is a modified adaptive sliding mode controller, where the control law is the same as \eqref{eq.u}, and the adaptation law \eqref{eq.gain} is modified by a small constant value to make the control gains decrease when the sliding variables are neighboring the sliding hyperplane, \emph{i.e.,}
\begin{IEEEeqnarray}{rCl}
	\dot{\Gamma}&=&\bar{\Gamma}\cdot\textup{diag}\{\vert s_1 \vert,\vert s_2 \vert,\vert s_y \vert\} \nonumber \\ &\:&\cdot\textup{diag}\{\textup{sgn}(\vert s_1 \vert-\epsilon),\textup{sgn}(\vert s_2 \vert-\epsilon),\textup{sgn}(\vert s_y \vert-\epsilon)\} \nonumber
\end{IEEEeqnarray}
where the overestimation of the control gains is mitigated by the modified adaptation law.

The other one is a traditional sliding mode controller with an iterative learning law. Notice that, the control gains are fixed and set as the same value with the initial control gains of the proposed method, \emph{i.e.,} $\Gamma=\Gamma(0)=1$. Besides, the iteration number is also 6, which is exactly the same as the iteration number in the experiment of the proposed method. For the sake of clarity, the comparative experiments are listed below:

Controller 1 (C1): Global iterative sliding mode controller.

Controller 2 (C2): Adaptive sliding mode controller.

Controller 3 (C3): Traditional sliding mode controller with iterative learning law.

Through this comparison, we would like to show that our proposed method can achieve better contouring performance with lower control gains and higher sliding accuracy (at least at the same levels and typically better and
improved).

Fig. \ref{fig.contour} shows the contouring performance in the cases of T1-C1, T1-C2, T1-C3, T2-C1, T2-C2, and T2-C3, respectively; where the colormap is used to illustrate and evaluate the contouring performance of each case. From the ticks of color bar, we can see that more `green' color represents better contouring performance while warmer color means a larger positive contouring error, and cooler color means a larger negative contouring error. Here, it can be obviously observed that the contouring performance of the proposed global iterative sliding mode control is better than the two comparative methods. Fig. 5 and Fig. 6 show the sliding variables, control gains, and control inputs of each carriage in Task 1 and Task 2 respectively. From these figures, compared with the modified adaptive sliding mode control method, the proposed method can achieve better contouring motion performance and the sliding variables are closer to the sliding hyperplane without control gain overestimation and larger input chattering. The reason is that the proposed method utilizes iterative learning law to make the process of control gain adaptation integrated closely into the iteration domain, so that the overestimation of control gain will be consequently mitigated. Also notice that, the red lines (C2) in the top three figures also demonstrate that the sliding variables get smaller (closer to the sliding hyperplane) in the time domain and they can achieve the similar values of the blue lines after long time adaptation while the control gains will be much higher than C1. These also show that the problem of control gains overestimation is solved. Compared with the traditional sliding mode controller with iterative learning law, the proposed method can also achieve better contouring performance. Moreover, the input chattering under the proposed method is lower than the comparative one and the sliding accuracy under the proposed method is higher. The reason is that the proposed method can ensure the establishment of sliding motion in both the time domain and the iteration domain globally. To show the results in terms of the tracking accuracy of each carriage, Table~\ref{tab:rmse} and Table~\ref{tab:maxae} illustrate the root-mean-square value and the maximum absolute value of the tracking errors between different control methods (in both Task 1 and Task 2). Similarly, Table~\ref{tab:rmssv} and Table~\ref{tab:maxasv} demonstrate the root-mean-square value and the maximum absolute value of the sliding variables. Furthermore, the convergence of contouring error and sliding variables is analyzed, where Fig. \ref{fig.converge_circle} and Fig. \ref{fig.converge_card} present the convergence behaviors in both contouring tasks. It is rather clear that RMSE, RMSSV-$s_1$, RMSSV-$s_2$, and RMSSV-$s_y$ decrease significantly after the first iteration; and in the following iterations, they are also decreasing, but with a smaller rate. From these figures, the convergence property and the establishment of sliding motion in the iteration domain are clearly validated.

\begin{table}[h]\centering
	\caption{RMSE of each carriage in the comparative experiments.}\ \label{tab:rmse}
	\begin{tabular}{|c|c|c|c|c|c|c|}
		\hline
		\multicolumn{7}{|c|}{RMSE ($\mu$m)} \\
		\hline
		\multirow{2}{*}{Carriage} & \multicolumn{3}{c|}{T1} & \multicolumn{3}{c|}{T2} \\ \cline{2-7} 
		& C1 & C2 & C3 & C1 & C2 & C3 \\ \hline
		$x_1$                     &    16.81         &   334.0   &    27.71    &     10.39        &  321.9   &  65.18      \\ \cline{1-7} 
		$x_2$                     & 11.62            &   300.0    &  18.27      &    10.05        &   218.4   &    40.77     \\ \cline{1-7} 
		$x_y$                     &    5.990         &  444.0    &   20.10       &     6.540        & 470.6     &  17.31       \\ \hline
	\end{tabular}
\end{table}

\begin{table}[h]\centering
	\caption{MaxAE of each carriage in the comparative experiments.}\ \label{tab:maxae}
	\begin{tabular}{|c|c|c|c|c|c|c|}
		\hline
		\multicolumn{7}{|c|}{MaxAE ($\mu$m)} \\
		\hline
		\multirow{2}{*}{Carriage} & \multicolumn{3}{c|}{T1} & \multicolumn{3}{c|}{T2} \\ \cline{2-7} 
		& C1 & C2 & C3 & C1 & C2 & C3 \\ \hline
		$x_1$                     &37.77             &  562.6   &   69.84        &35.34             &    679.4   &  121.1      \\ \cline{1-7} 
		$x_2$                     &   26.63          & 556.5   &     39.58      &   40.04          &386.0   &  100.2         \\ \cline{1-7} 
		$x_y$                     &16.80            &  708.7  &  156.3        &     17.23        &740.8     &  50.70      \\ \hline
	\end{tabular}
\end{table}

\begin{table}\centering
	\caption{RMSSV in the comparative experiments.}\ \label{tab:rmssv}
	\begin{tabular}{|c|c|c|c|c|c|c|}
		\hline 
		\multicolumn{7}{|c|}{RMSSV ($\times 10^{-4}$)}  \\ \cline{1-7}
		\multirow{2}{*}{Carriage} & \multicolumn{3}{c|}{T1} & \multicolumn{3}{c|}{T2} \\ \cline{2-7} 
		& C1         & C2   & C3      & C1         & C2   & C3      \\ \hline
		$x_1$                     & 3.25          & 32.32    &  4.41    & 2.62          & 23.27    &  5.50    \\ \cline{1-7} 
		$x_2$                     & 4.41          &26.43   &  4.38      & 2.31          &  17.82  &   4.15    \\ \cline{1-7} 
		$x_y$                     & 5.58        &    46.13  &   7.16    & 3.38          & 40.74   &  4.61     \\ \hline
	\end{tabular}
\end{table}

\begin{table}\centering
	\caption{MaxASV in the comparative experiments.}\ \label{tab:maxasv}
	\begin{tabular}{|c|c|c|c|c|c|c|}
		\hline 
		\multicolumn{7}{|c|}{MaxASV ($\times 10^{-4}$)}  \\ \cline{1-7}
		\multirow{2}{*}{Carriage} & \multicolumn{3}{c|}{T1} & \multicolumn{3}{c|}{T2} \\ \cline{2-7} 
		& C1         & C2   & C3      & C1         & C2   & C3      \\ \hline
		$x_1$                     & 15.36          & 85.49   &   15.58      & 16.64          & 58.94   &  24.38     \\ \cline{1-7} 
		$x_2$                     & 25.75          & 60.21    &  26.93     & 12.93          & 45.50   &   18.35     \\ \cline{1-7} 
		$x_y$                     & 13.30          & 177.8    &  29.16     & 43.03          & 120.6    &  35.63    \\ \hline
	\end{tabular}
\end{table}

\begin{remark}  \label{noise}
It is pertinent to mention that in our additional related efforts at validating and testing, we have also worked with the experimental case of additional exogenous disturbances (including, hence, instances of measurement noise). From the results of these experiments (which are available in reports, but which we are unfortunately unable to incorporate here due to strict page constraints), it is observed that the performance of the case with additional exogenous disturbance is at the same level to the case without additional exogenous disturbance. Therefore, we certainly can conclude that instances of these exogenous disturbance are well suppressed in our proposed method, and that the robustness against exogenous disturbance (including, hence, instances of measurement noise) can be validated successfully.
\end{remark}

\section{Conclusion}

In this work, an intelligent model-free control methodology is presented for contouring tasks of a flexure-linked biaxial gantry system. For such a multi-axis system, the coupling effects, along with the friction, nonlinearity, and disturbance are all considered as the uncertain dynamics; and a new formulation of a global iterative sliding mode control method is proposed in this paper to address this challenging problem. The method here utilizes a 2-DOF control structure, and is composed of an adaptive sliding mode controller with an incremental cascade iterative learning law. With the methodology developed here, the uncertainties arising from matched unknown dynamics in the time domain are effectively compensated for (via the first part of the composition involving the adaptive sliding mode control); and the undesirable effects arising from the remaining repetitive matched and unmatched uncertainties in the iteration domain are also additionally suppressed (via the second part of the composition involving the iterative learning law). Rather importantly too, with the approach utilized here, the chattering induced from the sliding mode control is also suitably suppressed during the iterations. All these improved outcomes (as stated in the work here) are supported by the pertinent proofs on the stability analysis in the time domain, and the convergence analysis in the iteration domain; as well as several real-time motion control experiments with different reference trajectories on the flexure-linked biaxial gantry system. This work thus certainly also points to promising further developments, with innovations using a model-free framework, to further improve the performance of industrial control systems. In the future, one of the directions of appropriate interest is to find an optimal learning function which can offer perhaps a possibly fastest convergence rate for the iterative learning law.

\section*{Appendix A}

Various matrices in \eqref{eq.gantry} are defined as 
\begin{equation}
	M = \begin{bmatrix}
		m_{11} & m_{12} & -m_e\textup{sin}(\Theta) \\
		m_{12} & m_{22} & 0 \\
		-m_e\textup{sin}(\Theta) & 0 & m_e
	\end{bmatrix}, \nonumber
\end{equation}
\begin{equation} 
	P = \begin{bmatrix}
		0 & p_{12} & -m_e\textup{cos}(\Theta)\dot{\Theta} \\
		p_{12} & p_{22} & 2m_eY\dot{\Theta} \\
		0 & -2m_eY\dot{\Theta} & 0
	\end{bmatrix}, \nonumber
\end{equation}
\begin{equation} 
	W = \begin{bmatrix}
		\mu_{k1}+\mu_{k2} & d_{12} & 0  \\
		d_{12} & d_{22} & 0 \\
		0 & 0 & \mu_{kY} \\
	\end{bmatrix},  \nonumber
\end{equation}
\begin{equation} 
	K = \textup{diag}\{0,k_{\tau 1}+k_{\tau 2},0 \}  \nonumber,
\end{equation}
\noindent where
\begin{IEEEeqnarray}{rcl}
m_{11} &=& m_e+m_{ca}+m_1+m_2, \nonumber \\
m_{12} &=& \frac{L_{ca}}{2}(m_1-m_2)\textup{cos}(\Theta)-m_eY\textup{cos}(\Theta), \nonumber\\
m_{22} &=& J_e+J_{ca}+m_eY^2+\bigg(\frac{L_{ca}}{2}\bigg)^2(m_1+m_2)\textup{cos}^2(\Theta), \nonumber\\
c_{12} &=& \bigg(-\frac{L_{ca}(m_1-m_2)}{2}+m_eY\bigg)\textup{sin}(\Theta)\dot{\Theta}-c_c\dot{\Theta}\nonumber\\ &&-m_e\textup{cos}(\Theta)\dot{Y},  \nonumber
\end{IEEEeqnarray}
\begin{IEEEeqnarray}{rcl}
c_{21} &=&c_c\dot{\Theta}, \nonumber\\
c_{22} &=& c_c\dot{X}+2m_eY\dot{Y}-\bigg(\frac{L_{ca}}{2}\bigg)^2(m_1+m_2)\textup{cos}(\Theta)\textup{sin}(\Theta)\dot{\Theta}, \nonumber\\
c_c &\fallingdotseq& \frac{m_e}{12}\bigg(X+\frac{L_e\textup{tanh}(\upsilon X)}{2}\bigg), \nonumber \\
d_{12} &=& \frac{L_{ca}}{2}(\mu_{k1}+\mu_{k2})\textup{cos}(\Theta), \nonumber\\
d_{22} &=&\mu_{\tau 1}+\mu_{\tau 2}+\bigg(\frac{L_{ca}}{2}\bigg)^2(\mu_{k1}+\mu_{k2})\textup{cos}(\Theta) \nonumber,\\
J_{ca} &=& \frac{m_{ca}}{12}(L_{ca}^2+W_{ca}^2), \nonumber \\
J_e &=& \frac{m_e}{12}\bigg(\big(\vert X\vert+\frac{L_e}{2}\big)^2+L_e^2 \bigg)+m_eY^2. \nonumber
\end{IEEEeqnarray}

\section*{Appendix B}

The contouring functions of the tasks in Section IV are shown below.

Task 1 (T1): Circle contour
\begin{IEEEeqnarray}{rcl}
	f(x_d,y_d)=x_d^2+(y_d+0.02)^2-0.02^2=0 \nonumber
\end{IEEEeqnarray}

Task 2 (T2): Cardioid contour
\begin{IEEEeqnarray}{rcl}
	f(x_d,y_d)= \Omega^2-\Xi^2(1-\nu)=0 \nonumber
\end{IEEEeqnarray}
where $\Omega = y_d+b(1-\nu)+d(2(1-2\nu)^2-1)-\rho$, $\Xi = a+2c-4c(1-\nu)$, $\nu = (\frac{x_d}{\psi})^{\frac{2}{3}}$,
with $a=0.065$, $b=0.025$, $c=0.01$, $d=0.005$, $\rho=0.025$, and $\psi=0.06$.

\bibliographystyle{IEEEtran}
\bibliography{IEEEabrv,Reference}

\begin{thebibliography}{10}
\providecommand{\url}[1]{#1}
\csname url@samestyle\endcsname
\providecommand{\newblock}{\relax}
\providecommand{\bibinfo}[2]{#2}
\providecommand{\BIBentrySTDinterwordspacing}{\spaceskip=0pt\relax}
\providecommand{\BIBentryALTinterwordstretchfactor}{4}
\providecommand{\BIBentryALTinterwordspacing}{\spaceskip=\fontdimen2\font plus
\BIBentryALTinterwordstretchfactor\fontdimen3\font minus
  \fontdimen4\font\relax}
\providecommand{\BIBforeignlanguage}[2]{{%
\expandafter\ifx\csname l@#1\endcsname\relax
\typeout{** WARNING: IEEEtran.bst: No hyphenation pattern has been}%
\typeout{** loaded for the language `#1'. Using the pattern for}%
\typeout{** the default language instead.}%
\else
\language=\csname l@#1\endcsname
\fi
#2}}
\providecommand{\BIBdecl}{\relax}
\BIBdecl

\bibitem{wang2020prediction}
Z.~Wang, C.~Hu, Z.~Yu, and L.~Zhu, ``Prediction model based contouring error
  iterative pre-compensation scheme for precision multi-axis motion systems,''
  \emph{IEEE/ASME Transactions on Mechatronics}, 2020.

\bibitem{yuan2016time}
M.~Yuan, Z.~Chen, B.~Yao, and X.~Zhu, ``Time optimal contouring control of
  industrial biaxial gantry: A highly efficient analytical solution of
  trajectory planning,'' \emph{IEEE/ASME Transactions on Mechatronics},
  vol.~22, no.~1, pp. 247--257, 2016.

\bibitem{mobayen2017composite1}
S.~Mobayen and F.~Tchier, ``Composite nonlinear feedback control technique for
  master/slave synchronization of nonlinear systems,'' \emph{Nonlinear
  Dynamics}, vol.~87, no.~3, pp. 1731--1747, 2017.

\bibitem{mobayen2018synchronization2}
------, ``Synchronization of a class of uncertain chaotic systems with
  lipschitz nonlinearities using state-feedback control design: A matrix
  inequality approach,'' \emph{Asian Journal of Control}, vol.~20, no.~1, pp.
  71--85, 2018.

\bibitem{hu2020gru}
C.~Hu, T.~Ou, Y.~Zhu, and L.~Zhu, ``{GRU}-type {LARC} strategy for precision
  motion control with accurate tracking error prediction,'' \emph{IEEE
  Transactions on Industrial Electronics}, vol.~68, no.~1, pp. 812--820, 2020.

\bibitem{ma2017novel}
J.~Ma, S.-L. Chen, N.~Kamaldin, C.~S. Teo, A.~Tay, A.~Al~Mamun, and K.~K. Tan,
  ``A novel constrained ${H}_2$ optimization algorithm for mechatronics design
  in flexure-linked biaxial gantry,'' \emph{ISA {T}ransactions}, vol.~71, pp.
  467--479, 2017.

\bibitem{xu2013design}
Q.~Xu, ``Design and development of a compact flexure-based {XY} precision
  positioning system with centimeter range,'' \emph{IEEE Transactions on
  Industrial Electronics}, vol.~61, no.~2, pp. 893--903, 2013.

\bibitem{zhu2016tmech}
H.~Zhu, C.~K. Pang, and T.~J. Teo, ``Integrated servo-mechanical design of a
  fine stage for a coarse/fine dual-stage positioning system,'' \emph{IEEE/ASME
  Transactions on Mechatronics}, vol.~21, no.~1, pp. 329--338, 2016.

\bibitem{wu2018design}
Z.~Wu and Q.~Xu, ``Design, fabrication, and testing of a new compact
  piezo-driven flexure stage for vertical micro/nanopositioning,'' \emph{IEEE
  Transactions on Automation Science and Engineering}, vol.~16, no.~2, pp.
  908--918, 2018.

\bibitem{kang2020six}
S.~Kang, M.~G. Lee, and Y.-M. Choi, ``Six degrees-of-freedom direct-driven
  nanopositioning stage using crab-leg flexures,'' \emph{IEEE/ASME Transactions
  on Mechatronics}, vol.~25, no.~2, pp. 513--525, 2020.

\bibitem{ma2017integrated}
J.~Ma, S.-L. Chen, N.~Kamaldin, C.~S. Teo, A.~Tay, A.~Al~Mamun, and K.~K. Tan,
  ``Integrated mechatronic design in the flexure-linked dual-drive gantry by
  constrained linear--quadratic optimization,'' \emph{IEEE Transactions on
  Industrial Electronics}, vol.~65, no.~3, pp. 2408--2418, 2017.

\bibitem{kamaldin2018novel}
N.~Kamaldin, S.-L. Chen, C.~S. Teo, W.~Lin, and K.~K. Tan, ``A novel adaptive
  jerk control with application to large workspace tracking on a flexure-linked
  dual-drive gantry,'' \emph{IEEE Transactions on Industrial Electronics},
  vol.~66, no.~7, pp. 5353--5363, 2018.

\bibitem{yang2010novel}
J.~Yang and Z.~Li, ``A novel contour error estimation for position loop-based
  cross-coupled control,'' \emph{IEEE/ASME Transactions on Mechatronics},
  vol.~16, no.~4, pp. 643--655, 2010.

\bibitem{yepes2013evaluation}
A.~G. Yepes, A.~Vidal, O.~L{\'o}pez, and J.~Doval-Gandoy, ``Evaluation of
  techniques for cross-coupling decoupling between orthogonal axes in double
  synchronous reference frame current control,'' \emph{IEEE Transactions on
  Industrial Electronics}, vol.~61, no.~7, pp. 3527--3531, 2013.

\bibitem{ma2018robust}
J.~Ma, S.-L. Chen, W.~Liang, C.~S. Teo, A.~Tay, A.~Al~Mamun, and K.~K. Tan,
  ``Robust decentralized controller synthesis in flexure-linked {H}-gantry by
  iterative linear programming,'' \emph{IEEE Transactions on Industrial
  Informatics}, vol.~15, no.~3, pp. 1698--1708, 2018.

\bibitem{ma2019parameter}
J.~Ma, S.-L. Chen, C.~S. Teo, A.~Tay, A.~Al~Mamun, and K.~K. Tan, ``Parameter
  space optimization towards integrated mechatronic design for uncertain
  systems with generalized feedback constraints,'' \emph{Automatica}, vol. 105,
  pp. 149--158, 2019.

\bibitem{li2017data}
X.~Li, S.-L. Chen, C.~S. Teo, and K.~K. Tan, ``Data-based tuning of
  reduced-order inverse model in both disturbance observer and feedforward with
  application to tray indexing,'' \emph{IEEE Transactions on Industrial
  Electronics}, vol.~64, no.~7, pp. 5492--5501, 2017.

\bibitem{li2020data}
X.~Li, H.~Zhu, J.~Ma, T.~J. Teo, C.~S. Teo, M.~Tomizuka, and T.~H. Lee,
  ``Data-driven multiobjective controller optimization for a magnetically
  levitated nanopositioning system,'' \emph{IEEE/ASME Transactions on
  Mechatronics}, vol.~25, no.~4, pp. 1961--1970, 2020.

\bibitem{yao2011orthogonal}
B.~Yao, C.~Hu, and Q.~Wang, ``An orthogonal global task coordinate frame for
  contouring control of biaxial systems,'' \emph{IEEE/ASME Transactions on
  Mechatronics}, vol.~17, no.~4, pp. 622--634, 2011.

\bibitem{hu2011global}
C.~Hu, B.~Yao, and Q.~Wang, ``Global task coordinate frame-based contouring
  control of linear-motor-driven biaxial systems with accurate parameter
  estimations,'' \emph{IEEE Transactions on Industrial Electronics}, vol.~58,
  no.~11, pp. 5195--5205, 2011.

\bibitem{wang2017newton}
Z.~Wang, C.~Hu, Y.~Zhu, S.~He, M.~Zhang, and H.~Mu, ``Newton-{ILC} contouring
  error estimation and coordinated motion control for precision multiaxis
  systems with comparative experiments,'' \emph{IEEE Transactions on Industrial
  Electronics}, vol.~65, no.~2, pp. 1470--1480, 2017.

\bibitem{chen2014mu}
Z.~Chen, B.~Yao, and Q.~Wang, ``{$\mu$}-synthesis-based adaptive robust control
  of linear motor driven stages with high-frequency dynamics: A case study,''
  \emph{IEEE/ASME Transactions on Mechatronics}, vol.~20, no.~3, pp.
  1482--1490, 2014.

\bibitem{hu2016advanced}
C.~Hu, Z.~Hu, Y.~Zhu, and Z.~Wang, ``Advanced {GTCF-LARC} contouring motion
  controller design for an industrial {X-Y} linear motor stage with
  experimental investigation,'' \emph{IEEE Transactions on Industrial
  Electronics}, vol.~64, no.~4, pp. 3308--3318, 2017.

\bibitem{han2020time}
S.~Han, H.~Wang, Y.~Tian, and N.~Christov, ``Time-delay estimation based
  computed torque control with robust adaptive {RBF} neural network compensator
  for a rehabilitation exoskeleton,'' \emph{ISA Transactions}, vol.~97, pp.
  171--181, 2020.

\bibitem{hu2020precision}
J.~Hu, C.~Li, Z.~Chen, and B.~Yao, ``Precision motion control of a 6-{D}o{F}s
  industrial robot with accurate payload estimation,'' \emph{IEEE/ASME
  Transactions on Mechatronics}, vol.~25, no.~4, pp. 1821--1829, 2020.

\bibitem{chen2020optimal}
X.~Chen, H.~Zhao, H.~Sun, S.~Zhen, and A.~Al~Mamun, ``Optimal adaptive robust
  control based on cooperative game theory for a class of fuzzy underactuated
  mechanical systems,'' \emph{IEEE Transactions on Cybernetics}, 2020.

\bibitem{edwards1998sliding}
C.~Edwards and S.~Spurgeon, \emph{Sliding Mode Control: Theory and
  Applications}.\hskip 1em plus 0.5em minus 0.4em\relax CRC Press, 1998.

\bibitem{xu2011micro}
Q.~Xu and Y.~Li, ``Micro-/nanopositioning using model predictive output
  integral discrete sliding mode control,'' \emph{IEEE Transactions on
  Industrial Electronics}, vol.~59, no.~2, pp. 1161--1170, 2011.

\bibitem{xu2012identification}
Q.~Xu, ``Identification and compensation of piezoelectric hysteresis without
  modeling hysteresis inverse,'' \emph{IEEE Transactions on Industrial
  Electronics}, vol.~60, no.~9, pp. 3927--3937, 2012.

\bibitem{li2021simultaneous}
D.~Li, S.~S. Ge, and T.~H. Lee, ``Simultaneous-arrival-to-origin convergence:
  Sliding-mode control through the norm-normalized sign function,'' \emph{IEEE
  Transactions on Automatic Control}, 2021.

\bibitem{mobayen2016novel4}
S.~Mobayen, ``A novel global sliding mode control based on exponential reaching
  law for a class of underactuated systems with external disturbances,''
  \emph{Journal of Computational and Nonlinear Dynamics}, vol.~11, no.~2, 2016.

\bibitem{mobayen2018adaptive3}
------, ``Adaptive global terminal sliding mode control scheme with improved
  dynamic surface for uncertain nonlinear systems,'' \emph{International
  Journal of Control, Automation and Systems}, vol.~16, no.~4, pp. 1692--1700,
  2018.

\bibitem{plestan2010new}
F.~Plestan, Y.~Shtessel, V.~Bregeault, and A.~Poznyak, ``New methodologies for
  adaptive sliding mode control,'' \emph{International Journal of Control},
  vol.~83, no.~9, pp. 1907--1919, 2010.

\bibitem{lee2007chattering}
H.~Lee and V.~I. Utkin, ``Chattering suppression methods in sliding mode
  control systems,'' \emph{Annual Reviews in Control}, vol.~31, no.~2, pp.
  179--188, 2007.

\bibitem{huang2008adaptive}
Y.-J. Huang, T.-C. Kuo, and S.-H. Chang, ``Adaptive sliding-mode control for
  nonlinear systems with uncertain parameters,'' \emph{IEEE Transactions on
  Systems, Man, and Cybernetics, Part B (Cybernetics)}, vol.~38, no.~2, pp.
  534--539, 2008.

\bibitem{gonzalez2011variable}
T.~Gonzalez, J.~A. Moreno, and L.~Fridman, ``Variable gain super-twisting
  sliding mode control,'' \emph{IEEE Transactions on Automatic Control},
  vol.~57, no.~8, pp. 2100--2105, 2011.

\bibitem{shtessel2012novel}
Y.~Shtessel, M.~Taleb, and F.~Plestan, ``A novel adaptive-gain supertwisting
  sliding mode controller: Methodology and application,'' \emph{Automatica},
  vol.~48, no.~5, pp. 759--769, 2012.

\bibitem{wang2020iterative}
W.~Wang, J.~Ma, X.~Li, Z.~Cheng, H.~Zhu, C.~S. Teo, and T.~H. Lee, ``Iterative
  super-twisting sliding mode control for tray indexing system with unknown
  dynamics,'' \emph{IEEE Transactions on Industrial Electronics}, vol.~68,
  no.~10, pp. 9855--9865, 2020.

\bibitem{bristow2006survey}
D.~A. Bristow, M.~Tharayil, and A.~G. Alleyne, ``A survey of iterative learning
  control,'' \emph{IEEE Control Systems}, vol.~26, no.~3, pp. 96--114, 2006.

\end{thebibliography}

\end{document}